**Title:** Estimating beneficiaries of the child tax credit: past, present, and future

**Authors:** Ashley Nunes[1,2], Chung Yi See[1,3], Lucas Woodley[1,3], Nicole A. Divers[1], and Audrey L. Cui[1,4]

[1] R Street Institute
Washington, DC 20005

[2] Labor and Worklife Program
Harvard Law School
Cambridge, MA, 02138, USA

[3] Department of Economics
Harvard College
Cambridge, MA, 02139, USA

[4] Haas School of Business
University of California at Berkeley
Berkeley, CA, 94720, USA

**Corresponding Author:** Ashley Nunes, anunes@rstreet.org

**Abstract:** Government efforts to address child poverty commonly encompass economic assistance programs that bolster household income. The Child Tax Credit (CTC) is the most prominent example of this. Introduced by the United States Congress in 1997, the program endeavors to help working parents via income stabilization. Has this aim been achieved? If so, to what extent, and by whom? We exploit a rich data set spanning 15-years to address this question. Our analysis—which documents clear, consistent, and compelling evidence of gender inequity in benefits realization—yields four key findings. First, stringent requisite income thresholds disproportionally disadvantage single mothers, a reflection of the high concentration of this demographic in lower segments of the income distribution. Second, married parents—and, to a lesser extent, single fathers—are the primary beneficiaries of the CTC program when benefits are structured as credits rather than refunds. Third, making program benefits more generous disproportionally reduces how many single mothers—relative to married parents and single fathers—can claim this benefit. Fourth and finally, increasing credit refundability can mitigate gender differences in relief eligibility, although doing so imposes externalities of its own. Our findings can inform public policy discourse surrounding the efficacy of programs like the CTC and the effectiveness of programs aimed at alleviating child poverty.

## 1.0 Introduction

Who benefits from the Child Tax Credit (CTC)? The CTC is one of the largest safety net programs in the United States. Introduced with bipartisan support in 1997, the program reflected political recognition that elements of the federal tax code—specifically nominally fixed dependent exemptions—did not "reduce tax liability by enough to reflect a family's reduced ability to pay taxes as family size increase(d)" (1). The CTC's intent was to address this by offering financial relief to middle and upper-income individuals and families with children.

Since the CTC's introduction, various laws have modified key parameters of the program, expanding access to more working families and increasing the magnitude of benefits (2). What was structured in 1997, as a $500-per-child nonrefundable credit has, as of 2018, become a $2,000-per-child benefit. Like other tax credits, the CTC reduces tax liability by the amount of the credit. Moreover, a portion of CTC, known as the Additional Child Tax Credit (ACTC), is 'refundable.' Should a parent's annual earned income (hereafter referred to as '*income*') make them ineligible to claim the full $2,000 benefit (because their tax liability falls short of this amount), these individuals may still realize some relief, albeit as a refund rather than a credit.

The amount of relief realized under these circumstances is dictated by program design features. These features include a refundability threshold and refundability rate. The former reflects the minimum income threshold ($2,500 in 2018) required to claim any benefit, the latter, how much of a refund ($1,400 in 2018) a household may realize above the refundability threshold. The CTC and the ACTC can, under certain conditions, be claimed simultaneously although the total benefit realized cannot exceed the stipulated per-child benefit ($2,000 in 2018). The CTC also has a phaseout mechanism that delivers gradual reductions in benefits as income approaches specific thresholds. In 2018, benefits decreased by $50 for every $1,000 single parents and married parents earned over $200,000 and $400,000 respectively.

The intricacies of the CTC program reflect two concurrent realities, the first being that income scarcity—particularly in families with children—warrants attention (3, 4). Higher income supports reductions in family stress that subsequently promote healthy parenting. It also allows parents to invest in measures that advance physical and mental functions in their children. The result is superior short and long-term development coupled with improved well-being (5, 6). Consequently, facilitating these outcomes is important. It is particularly important in the United States which, despite its sizable wealth, exhibits high rates of child poverty relative to other high-income countries (7-11).

However, wealth alone does not eliminate the need for prudence, financial or otherwise. The cost of child poverty alleviation programs in the United States ($118 billion in 2018) has drawn criticism, as has whether, and to what extent, such programs demonstrably alleviate child poverty (12-14). Broader fears persist that, when left unchecked, programs like the CTC can increase the number of households that rely primarily on government benefits as a substitute for, rather than a complement to, employment (6, 13, 15-17). Evidence and debate surrounding the labor market impact of unconditional cash transfers to parents is voluminous, compelling, and contradictory (18-20).

The CTC's design addresses some of these concerns by tying the receipt of benefits to employment. Unlike other countries (e.g., Canada, Germany, and Finland) that offer an unconditional child allowance to parents, the complex interaction of refundability thresholds, refundability rates, phaseout thresholds, and limits on the magnitude of the CTC and ACTC dictate specific parental labor force participation standards to access program benefits. As evidenced by higher income, greater participation yields greater realization of these benefits.

Who realizes these benefits and to what extent? How does the inclusion of or changes to design features of the CTC program impact parental ability to realize program benefits (hereafter referred to as relief eligibility)? Do some features levy more influence than others? If so, which ones and on whom?

Our work addresses these questions. To do so, we source data from the United States Census Bureau and the Bureau of Labor Statistics. Our data set spans 15 years (2003 - 2018), a timeframe associated with key changes to the CTC program. We leverage this data to estimate how the inclusion of and changes to CTC program design features impact relief eligibility. We further assess the extent to which relief eligibility has been equitably distributed, focusing on marital status (married or single) and gender (male or female).

Our work differs from previous efforts in several ways. First, we examine relief eligibility for the CTC program, not the short and long-run impact of participation. Consequently, we do not address, among other things, whether participation in the CTC program impacts labor market supply (6, 13, 16-17, 21-25). Nor do we estimate whether the long-run benefits that may accrue to recipients and society from investing in the CTC program outweigh the short-term costs incurred by taxpayers (12, 13, 26-28). While discussion of these issues is undoubtedly timely, should be debated, and warrants further scrutiny, it is not the focus of our efforts.

Like others, we simulate the impact of CTC changes under reforms that amend one or more rule(s) (29). Like others, we examine the distributional effects of these changes (29, 30). However, unlike previous work, our focus is on identifying which parental groups—stratified by gender—qualify for relief and what type of relief they qualify for owing to rule changes (31, 32). Unlike previous work, we also scrutinize how individual changes to CTC design features impact relief eligibility. This approach offers more precision when estimating the impact of CTC design features. It also reflects the governance challenges associated with limited capital.

Most importantly, we distinguish our work from previous efforts by relating long-run programmatic changes to the CTC program (e.g., adjustments to the refundability threshold, refundability rate, and credit size) to relief eligibility. Our efforts, which entail consideration of evolving federal tax code, CTC program parameters, and demography, represent, to our knowledge, the first of their kind and afford greater precision in estimating how specific attributes of the CTC program (rather than the program-at-large) have impacted relief eligibility over time. This facilitates the identification of what program parameters may—moving forward—impact program access.

Our efforts can inform ongoing public policy debate about child poverty alleviation efforts. The American Rescue Plan (ARP) Act of 2021 provided $1.9 trillion in transfer payments, new spending and other assistance to households, businesses, and public institutions. Among the most significant of these was a temporary expansion in the amount of and eligibility for the CTC. However, this program's future remains uncertain, owing in part to program costs. Our results help address this concern by providing evidence-based guidance on what policy levers are most effective in increasing relief eligibility for the program, compared to a one-size-fits-all approach.

**2.0 Method**

Our analysis uses a three-step approach. First, we extract year-on-year annual income for parental groups given that income is a prerequisite for relief eligibility. Concurrently, we approximate the number of children associated with these groups. Second, using federal tax guidelines, we calculate the income thresholds that dictate relief eligibility leveraging the aforementioned approximations (Fig. 1). Third, we develop a data-informed model to determine how many married parents, single fathers, and single mothers qualify for various forms of CTC relief. We subsequently isolate the individual contributions of CTC design features.

*2.1 Approach Specification*

Step 1: Parental Income/Dependent Estimation: We first retrieve data from the U.S. Census Bureau Current Population Survey (CPS) to estimate the total number of married parents, single fathers, and single mothers, broken down by income level. The CPS's Annual Social and Economic Supplement (ASEC) provides demographic, economic, and social characteristics of the population based on a survey of more than 75,000 households. To produce national estimates from survey data, a statistical weight for each person in the sample is developed from a series of weighting steps involving benchmarking CPS estimates of the population to independent estimates of the current population (33).

The CPS asks detailed questions covering social and economic characteristics, such as the number of related children in the respondent's household, respondent marital status, and respondent income. This information can be queried using the CPS Table Creator to generate customized data tables from the ASEC for a specified time frame (34, 35). Using this feature, we retrieve, year-on-year, the number of Adult Civilian Workers, broken down by marital status and gender, who had at least one child under 18 in the household. Adult Civilian Workers include all workers aged 18 or older who are part of the non-institutionalized US population. This population is of the age most likely to be affected by the CTC and closely approximates the total number of parents in the workforce. This is further subdivided by household income level, in $2,500 increments, from $0 to $99,999 annual pre-tax-and-transfer income. The result is an estimation of the total number of married parents, single fathers, and single mothers.

We also use the CPS to estimate the number of children (aged under 18) associated with each parental group. Estimating the number of so-called 'dependents' is necessary as prior to 2018, dependent quantity influenced the magnitude of tax relief that taxpayers, regardless of marital status, gender, or income, were eligible for. Offered via personal/dependent exemptions for parents and children, such relief constitutes a 'below the line' deduction—in addition to the usual standard deduction—that was subtracted from income estimates, which subsequently impacted income thresholds required to realize full CTC benefits. Consequently, our model accounts for this parameter.

Step 2: Requisite Income Threshold Approximation: In this step, using tax guidelines from the Internal Revenue Service, we calculate the income thresholds required of married and single-parent households to qualify for CTC relief. These thresholds are classified as followed: 1) income below which no relief is realized, 2) income required to earn some ACTC, 3) income required to earn the full ACTC, 4) income required to earn the full CTC, 5) income required to earn some CTC and 6) income above which no relief is realized. This classification mirrors specific relief eligibility categories; namely, (a) ineligible for any relief (owing to low income), (b) eligible for some ACTC relief; (c) eligible for full ACTC relief; (d) eligible for full CTC relief, (e) eligible for some CTC relief (owing to high income) and (f) ineligible for any relief (owing to high income).

Our serial listing of relief eligibility categories reflects the impact of income increases within a household. When income is below the refundability threshold, no relief is realized (category a). As income rises above the threshold, partial relief is realized via the ACTC (category b). Added income prompts realization of the full ACTC (category c) and subsequently, the full CTC (category d). However, continued income increases produce reductions in the magnitude of CTC realized (category e) owing to the phaseout threshold. Once income exceeds the total phaseout threshold (category f), no relief is realized.

Our income threshold estimates account for, year-on-year, the number of children in a household, the household's tax filing status, and CTC program parameters. The average number of children in a household is approximated for married parents, single fathers, and single mothers. A household's tax filing status assumes that married parents will file a joint return and single parents will opt for head of household status as these choices afford more favorable taxation conditions (i.e., a larger standard deduction and requisite personal/dependent exemptions). CTC program parameters (e.g., refundability rates, refundability thresholds, and ACTC/CTC magnitude) are ascertained from government sources (2).

Step 3: Relief Eligibility Approximation: Having ascertained year-on-year income for married parents, single fathers, single mothers, the average number of children for each of these groups, and the requisite income thresholds these groups must meet to qualify for CTC relief, we now approximate how many married parents, single fathers, and single mothers meet the requisite relief standard. Our model assigns parents to one of six groups based on income thresholds for 2003 - 2018, which considers the requisite CTC relief criteria. For example, in 2009, the refundability threshold was $3,000, the refundability rate, 10 percent, and both the ACTC and CTC, $1,000. During this year, a single parent with one child required $9,667 and $25,650 in income to realize the full ACTC and CTC, respectively.

Consequently, for that year, our model assigns parents who earned more than $25,650 but less than the $75,000 phaseout threshold to the full CTC relief group. Parents who earned between $9,667 and $25,650 are assigned to the full ACTC relief group, parents who earned less than $9,667 but more than $3,000 (the refundability threshold) are assigned to the some ACTC group, and those earning less than $3,000, to the no relief group. Parents who exceed the phaseout threshold but not the total phaseout threshold ($75,000 and 95,000 respectively in 2009 for single parents) are assigned to the 'some' CTC relief group. Those whose income exceeds the total phaseout threshold are assigned to the no relief group (owing to high income). This process is replicated year-on-year across parental groups. The result is a comprehensive picture of how relief eligibility for the CTC program has changed over the timeframe.

*2.2 Vocabulary Specificity*

Our use of the term 'full ACTC' warrants further treatment. The term admittedly implies benefits realization solely as a refund given that the ACTC is – in existing policy discourse - viewed as being 'refundable.' However, owing to year-on-year heterogeneity in the refundability threshold and the federal tax code, this is not necessarily the case. For example, in 2003, the maximum envisioned per child benefit was $1,000, which was available via both the CTC and the ACTC. Realizing this benefit via the ACTC for single parents (assuming 1 child) required an income of $16,795 (Fig. 1; Appendix A1). However, at that income level, only $630 of the ACTC would be realized as a refund; the remaining balance ($270) would be delivered as a credit. Income greater than $16,795 (but less than the income threshold required for full CTC realization) would decrease the size of the refund while increasing the size of the credit (as the total benefit may not exceed $1,000) whereas income lower than $16,795 would lower both the refund and credit (eventually producing only a refund, assuming income still exceeds the refundability threshold).

The necessity of claiming a partial credit to realize a 'full' refund reflects in part, intricacies of the federal tax code which offers some tax free relief to all tax filers. Only income above a specified threshold (the standard deduction and personal exemptions combined pre 2017) are subject to taxation. Given that the refundability rate delivers relief as income rises above the refundability threshold, there are instances, specifically, when the refundability threshold lies in close proximity to the standard deduction (and personal exemptions), where realizing the maximum envisioned benefit via a refundable pathway necessitates producing some taxable income. Such was the case between 2003 and 2008 (Fig. 1). Consequently, throughout our paper, the term 'full ACTC' refers to instances where the maximum realizable ACTC benefit occurs *primarily* as a refund. We note that a similar issue does not persist with regard to full CTC relief, as this relief category depends solely on earning taxable income.

Another aspect of the CTC program warrants mention. Prior to 2018, the maximum envisioned per child benefit of$1,000 could be realized by claiming the full ACTC or the full CTC, this owing to parity between the ACTC and CTC. However, whereas in 2018, the maximum envisioned per child benefit increased to $2,000, only $1,400 of that benefit was realizable via the ACTC. Conversely, the full benefit was realizable via the CTC. Under this circumstance, claiming the maximum envisioned per-child benefit required either, a) accruing $2,000 in tax liability (thereby allowing the beneficiary to claim the full CTC), or b), claiming the full ACTC ($1,400) *and* accruing $600 in tax liability (which necessitates income above the standard deduction). Consequently, when discussing 2018 relief eligibility throughout our paper, the term 'full ACTC' should also not—unlike the term, 'full CTC'—be equated with realization of the maximum envisioned per child benefit ($2,000) but rather the maximum benefit associated with the ACTC (i.e., $1,400). We do, however, consider the alternative scenario (ACTC + CTC combined) as a separate benefits category in our analysis and present/discuss it where appropriate.

*2.3 Data Source Rationale*

Unlike previous work that uses individual-level CPS ASEC data to assess relief eligibility, our approach leverages the CPS Table Creator. Two factors motivate this choice. First, using the Table Creator allows assessment of how relative changes to program parameters (e.g., refundability threshold, phase in rates, etc.) impact relief eligibility. Secondly, individual-level income data is prone to some imprecision, owing to discrepancies between respondents' actual versus reported income (36). The reasons for these discrepancies are varied and include definitional issues, recall and salience problems, confusion, and sensitivity (37). Using the CPS Table Creator affords, with appropriate corrections,  mitigation of reporting imprecision by allocating individual-level income data to income bins. This approach (detailed in Section 2.4 below) allows consideration of more and less conservative estimates of relief eligibility given heterogeneity in the truthfulness of participant responses. Consistency across estimates would highlight robust effects that warrant consideration.

Consider a single parent with one child reporting $22,900 in annual income for 2003. The income threshold required to claim the full CTC was $23,100.Based on reported income, this parent would be ineligible to claim the full CTC. However, if reported income is an underestimate of that parent's true income, the parent would erroneously be classified as ineligible for full CTC relief. By relegating this parent to an income bin (in this case, the $22,500 - $24,999 bin), we account for both the case where $22,900 is their accurately reported income, whereby the parent does not qualify for full CTC, and the case in which the parent's income is under-reported. By omitting or including the $22,500 - $24,999 bin based on more or less conservative estimates of relief eligibility, both cases of accurate- and erroneously reported income are accounted for. Thus, by generating a more and less conservative estimate of relief eligibility, we can

better understand the extent to which relief eligibility could vary across parental groups and assess relief eligibility in a worst-case scenario through the conservative estimates.

Omitting individual-level CPS ASEC data admittedly precludes consideration of households where income exceeds $99,999 (as the CPS Table creator relegates these households to a single group). However, we argue that these households are less relevant to our analysis as their income level far exceeds the federal poverty threshold (38). Because the CTC is increasingly viewed as an income stabilization tool for low-income households (although we emphasize this was not the program's original intent), programmatic changes over time emphasize increasing accessibility to lower, rather than higher, income households (5, 6, 9-10, 12). As households whose annual income exceeds $99,999 fall in the latter category, they are arguably less central to efforts to alleviate child poverty via the CTC (8, 12, 17), and by consequence, our assessment of moderate-to-low income households' ability to realize this benefit.

*2.4 Upper and Middle Bound Estimates*

By considering heterogeneity in the number of children related to a household, and the number of people moving between adjacent relief eligibility buckets, year-on-year, we provide a middle-estimate of relief eligibility that more closely reflects the use of individual-level data and an upper-estimate that produces more conservative estimates of relief eligibility.

1. Children Estimation: The CPS Table Creator categorizes respondents by the number of related children in their household, but not the number of own children. Related children are presumed, under Census Bureau guidelines, to be aged under 18, and related to the survey respondent by birth, marriage, or adoption (39). This definition varies from 'own children,' considered to be sons and daughters, including stepchildren, and adopted children of the respondent. The use of 'related children' over 'own children' may erroneously increase the number of reported children per household. The result is a potential overestimate of the income threshold required to claim full CTC relief, as this threshold (compared to those associated with the ACTC) depends on the total standard deduction (which is impacted by the number of children). More children raise the total standard deduction, thereby increasing income thresholds required to claim full CTC relief.

To address this, we estimate an upper and middle bound for the number of children. Our upper bound is a more conservative estimate of the number of children within a household, which is used in our more conservative estimate of relief eligibility. This is set to the minimal number, namely one. We assume that at least one child in a household is a dependent (for tax purposes) of the respondent. Our middle-bound estimate produces more moderate estimates for the number of children, which is used in our middle-estimate of relief eligibility. Here, we consider CPS responses in their totality. Each respondent can indicate a discrete number between 0 and 7, or 8 or more, for the number of related children under 18. Thus, we calculate the total number of related children under 18 for married respondents, single fathers, and single mothers, year-on-year. This is done by multiplying the number of respondents by the number of children indicated and summing the product. For respondents indicating 8 or more children, the number of respondents is multiplied by 8. We subsequently calculate the average number of children in each parental group by dividing the total number of children in that group by the number of parents.[1]

---

[1] The 'under 18' query introduces an additional potential error source as the CTC is only available to households with children aged under 17 (2). Consequently, our model may slightly overestimate the total number of households that should be eligible for relief because it counts—by virtue of using CPS data—those parents of children aged between 17 and 18 as being relief eligible when they are in fact, not. However, we do not expect this discrepancy to

2. Income Bracket Estimation:  Income brackets specified by the CPS Table Creator overlap with income thresholds required for CTC relief. For example, a single parent with one child who earned $25,500 in 2009 must select the closest applicable income bracket on the CPS (in this case, $25,000 to $27,499). During the same year, the requisite income threshold for full CTC relief was $25,650. In such cases, based on our data sample, we cannot definitively ascertain whether this respondent was eligible for full CTC relief (in the example, the respondent was not). To address this, we calculate an upper and middle bound when estimating how many parents are eligible for relief.

To calculate the upper bound, groups in the income bucket containing the requisite income threshold are summed with the lower credit eligibility group and subsequently 'omitted' from a target category. Consider a case where we estimate how many parents qualify for full CTC relief. If the income required to claim the full CTC is $25,650, and the overlapping CPS income bracket is $25,000-$27,499, we only count respondents whose reported income exceeds $27,500 as being eligible for full CTC relief. Respondents who fall in the $25,000-$27,499 category are counted as being ineligible for full CTC relief. Instead, they are considered eligible for full ACTC relief and added to respondents who earn more than $9,666 (the income threshold required to realize full ACTC benefits) but less than $28,000. This approach produces more conservative estimates of parents eligible for various forms of CTC relief.

Our middle-bound estimate produces more moderate estimates of the number of parents eligible for CTC relief. Here, we consider where the requisite income threshold falls in the CPS income bracket. If the threshold falls at least halfway through the income bracket, the number of parents in that bracket is summed with the lower eligibility group. If, however, the threshold falls less than halfway into the bucket, the number of people in the bucket is summed with the target eligibility group. For example, if the income required to claim the full CTC is $27,000, assuming a CPS income bracket of $25,000-$27,499, those falling within this bracket are summed to those able to claim full ACTC (income above $9,667). If, however, the income required to claim the full CTC is $25,650, those falling in an income bracket of $25,000-$27,499 are considered eligible for full CTC relief.[2]

*2.5 Scenario Specification*

Consideration of heterogeneity in the number of children and accuracy of reported income produce an orthogonal set of scenarios (four in total). We restrict our presentation of results in this report to the most conservative and more moderate of these.

Our first scenario (hereafter referred to as S1) considers cases where parents, regardless of marital status or gender, claim one child as a dependent year-on-year, and reported household income is subjected to an upper bound cutoff. Our second scenario (hereafter referred to as S2) considers a case where the number of children claimed as dependents are drawn directly from CPS respondents and stratified by marital status and gender, year-on-year. For each year between 2003 - 2018, based on CPS data we

---

significantly impact our analysis as our analysis scrutinizes differences between parental groups. There is no evidence suggesting that any one parental group has a higher percentage of children aged between 17 and 18.

[2] The CPS queries respondents on overall income (which includes dividends, rent, royalties and employment related wages to name a few), rather than solely 'earned income' (i.e., employment related wages). The latter is the only applicable parameter when estimating CTC relief eligibility (2). Consequently, by virtue of considering overall income, our model may slightly overestimate the *total* number of households that should be eligible for relief. However, this discrepancy is unlikely to impact our findings as we scrutinize relief eligibility differences between parental groups. There is no evidence suggesting that any one group draws higher rates of overall earned income than others.

estimate the number of children for married parents, single fathers, and single mothers, separately and use that data to ascertain the total applicable standard deduction, and consequently requisite income thresholds for CTC relief. Income estimates in this scenario are, for relief eligibility purposes, subject to a middle-bound cutoff (described earlier).

*2.6 Analysis Overview*

To assess the impact of programmatic changes on the CTC, we employ comparative analyses and linear fixed effects regressions. Specifically, we compare eligibility across parental groups in 2003 versus 2017 to capture the effects of reductions in the refundability threshold. We also regress eligibility estimates on year, parental group, and an interaction term for each category of relief. We subsequently compare the ratios of parents qualifying for full CTC versus ACTC relief with and without numerical parity. Furthermore, we isolate the effects of individual parameter changes that occurred in 2018 using a piecemeal analysis. That is, we estimate the proportion of parents eligible for full CTC and ACTC relief given changes to the maximum CTC/ACTC, standard deduction and phaseout thresholds, and subsequently assess resulting inequities using a difference-in-differences analysis. A detailed overview of numerical parameters, assumptions, and model outputs is available in the supplementary information section.

**3.0 Results and Discussion**

Our results and discussion are structured as follows. In Section 3.1, we present key findings of our work. Relief eligibility between 2003 and 2017 is first assessed given the relative stability of program features during that period, followed by the impact of 2018 specific changes (Fig. 1). In Section 3.2, we discuss the equity implications of our findings. Across both sections, we underscore that treatment and discussion of relief eligibility estimates only apply to households whose income is less than $100,000. A summary of the generalized applicability of these estimates to the population at large is provided in the Appendix (A2) and indicated on relevant figures. We also note that across both sections, S1 model outputs are presented first, with S2 outputs presented alongside in parenthesis. The presentation of a single number indicates congruence between S1 and S2.

*3.1 Key Findings*

Our first finding is that whereas a stringent refundability threshold—which is the minimum income required to qualify for relief—limits program eligibility, relaxing this threshold yields heterogenous benefits distribution across parental groups. Compared to married parents and single fathers, single mothers have disproportionally benefited from refundability threshold reductions (Fig. 2a, 3a) which subsequently has increased their eligibility for some ACTC relief (Fig. 2b, 3b). In 2003, 11.06 (7.58) percent of single mothers were—given a $10,500 refundability threshold—ineligible for any form of CTC relief, compared to 2.35 (1.52) and 4.06 (2.90) percent of married parents and single fathers, respectively. In 2017, during which the refundability threshold was $3,000, 2.03 (1.07) percent of single mothers were ineligible for any form of CTC relief compared to 0.33 (0.16) and 0.53 (0.39) percent of married parents and single fathers respectively. We attribute these effects to the lower income that single mothers realize, which subsequently increases their dependence on refundability threshold reductions as a pathway towards relief realization.

A lower refundability threshold impacted single mothers in another way; it disproportionally increased their eligibility for full ACTC relief (Fig. 2c, 3c). The ACTC allows, within specified parameters, refunds of up to 15 percent of income exceeding the refundability threshold. Consequently, a lower refundability threshold, largely owing to the 2008 Emergency Economic Stabilization Act (EESA) and the 2009 American Recovery and Reinvestment Act (ARRA), lowered the income requirement to fully realize the ACTC. In 2017, 23.68 (33.16) percent of single mothers were—owing to the $3,000 refundability threshold—eligible to claim the full ACTC, compared to 14.57 (27.95) and 11.48 (18.09) percent of married parents and single fathers respectively. Were the refundability threshold in 2017 held at 2003 levels (i.e., $10,500), 17.65 (26.76) percent of single mothers would be eligible for full ACTC relief, as would 13.21 (20.97) and 10.53 (15.20) percent of married parents and single fathers, respectively.

Our second finding is that the income requirements for claiming the full CTC have disproportionally favored married parents and to a lesser extent single fathers (Fig. 2d, 3d). Between 2003 and 2017, a period during which the magnitude of the full CTC remained unchanged, 83.38 (71.31) percent and 64.29 (53.90) percent of married parents and single fathers respectively were eligible to claim the full CTC compared to 56.48 (43.19) percent of single mothers. A fixed effects linear regression shows this difference is statistically significant (See Supplementary Information, Section III for details). The stratification of full relief eligibility by marital status and gender reflects the lower income single mothers realize, and by consequence, the lower tax liability they incur, which decreases full CTC relief eligibility. Lower income among single mothers also explains why they are—compared to single fathers—less likely to realize some CTC relief (Fig. 2e, 3e), a category where earnings must be high enough to exceed the phaseout threshold (i.e., $75,000 between 2003 and 2017). Similar gender effects persist when income exceeds the total phaseout threshold (Fig. 2f, 3f).

However, we also find that single mothers' comparative inability (relative to married parents and single fathers) to qualify for full CTC relief (i.e., $1,000) has not entirely prevented their realization of monetary relief of similar magnitude, owing to the ACTC. Between 2003 and 2017, the ACTC offered identical benefits to the CTC ($1,000) albeit via a different pathway, i.e., as a refund versus a credit. As income thresholds for refund recovery are lower than credit recovery, more single mothers were eligible for this benefit. Consequently, some single mothers could, for a subset of years, realize the same magnitude of relief as married parents and single fathers despite earning less. Absent numerical parity between the CTC and ACTC (i.e., an ACTC of $1,000 and CTC of $2,000), our model estimates that 30.35 (40.72) percent fewer single mothers would—given homogeneity in the refundability threshold—be eligible to claim the same magnitude of relief between 2003 and 2017.

Income differentials between single mothers and married couples/single fathers also inform our third finding: single mothers are disproportionally impacted by increasing the magnitude of per-child benefits. Because relief realization in the CTC program dictates meeting specific income thresholds, increasing the magnitude of benefits makes their recovery more challenging. Realizing larger tax credits in particular necessitates incurring higher tax liability, which in turn necessitates earning more. Due to lower earnings, single mothers are more likely to be affected by a higher income requirement.

Consider 2018, a year during which the CTC increased in magnitude from $1,000 to $2,000. Were potential CTC beneficiaries in 2018 subjected to 2017 relief eligibility standards, our model estimates that 57.75 (47.71) of single mothers would be eligible to realize the full $1,000 credit as would 84.75 (76.67) and 60.52 (52.30) percent of married couples and single fathers, respectively (Table 1a). However, raising the credit to $2,000 would reduce eligibility for single mothers to 42.56 (30.92), compared to 74.37 (58.26) and 50.05 (37.28) percent of married couples and single fathers respectively.

Although adjusting individual parameters highlights single parents' reduced eligibility in 2018, a cursory overview of our findings may suggest otherwise (Fig. 2d, 3d).[3] This discrepancy—i.e., increased eligibility despite a higher requisite income threshold—is explained by the 2018 implementation of a higher phaseout threshold (and to a lesser extent, a lower standard deduction). A higher phaseout threshold precludes some potential beneficiaries from becoming relief ineligible owing to increased income, thereby raising the number of beneficiaries. However, here too gender discrepancies persist, as a difference-in-differences analysis shows 2018's programmatic changes benefitted single mothers significantly less than single fathers (see Supplementary Information, Section III for details). Single fathers in particular benefited from a higher phaseout threshold compared to single mothers. Owing largely to a higher phaseout threshold, full CTC relief eligibility for single mothers increased from 42.56 (30.92) to 58.44 (43.37) versus from 50.05 (37.28) to 76.72 (64.29) percent for single fathers.

Full ACTC relief eligibility in 2018 follows a similar trend. Whereas the 2018 ACTC increase from $1,000 to $1,400 disadvantaged all parental groups, single mothers were disproportionally affected (Table 1b). Yet, our analysis shows that full ACTC relief eligibility rose in 2018 compared to 2017 (Fig. 2c, 3c). This discrepancy is driven by the 2018 increase in the magnitude of the CTC. A higher CTC necessitates higher income for full relief eligibility, which precludes previously-eligible beneficiaries from full CTC relief.

---

[3] The impact of a higher phase out threshold on married parents are excluded from this portion of our analysis owing to insufficient data granularity. However, given the higher proportion of married parents in the upper segments of the income distribution, full CTC relief eligibility for this group would also likely rise.

However, these individuals can now claim the next available benefit, namely the full ACTC, which increases relief eligibility in this category.

*3.2 Addressing relief inequity*

Who benefits from the CTC program? The simple answer is more working parents than in years prior. Our model estimates that in 2018, 99.07 (99.47) percent of parents earning less than $100,000 were eligible for some form of CTC relief. Were parents subjected to 2003 relief eligibility standards, 95.88 (97.83) percent of them would qualify. We attribute this increase to a higher phaseout and, more relevant to poverty alleviation efforts, a lower refundability threshold. Single mothers have disproportionally benefited from the latter parameter, a reflection of their high concentration in the lower segments of the income distribution and, consequently, their reliance on low-income participation thresholds for program access.

However, whereas relief eligibility is one measure of relevance, another is what type of relief may be claimed and by whom. While relief eligibility carries an explicit income requirement, realizing the maximum envisioned per child benefit – when structured as a credit (rather than a refund) - carries an implicit marriage requirement. Year-on-year married parents and single fathers are better positioned to claim the full CTC compared to single mothers. Put simply, when it comes to realizing full CTC relief, it pays to be married, particularly if the beneficiary of the claim is female.

Gender differences in full CTC relief eligibility warrant particular scrutiny given heterogeneity in the average number of children. This heterogeneity impacts the income thresholds required to realize full CTC relief, with single mothers subjected to higher income requirements compared to single fathers (by virtue of claiming more dependents for tax relief). Considering this parameter yields more pronounced gender differences in full CTC relief eligibility. Between 2003 and 2017, 56.48 percent of single mothers could, on average – assuming one child as a dependent - claim full CTC relief compared to 64.29 percent of single fathers. However, after accounting for dependent heterogeneity, these figures fall to 43.54 percent and 54.44 percent respectively, a 3.09 percentage point increase in gender differences in relief eligibility (Fig.2d,3d).

Historical parity between the ACTC and CTC admittedly implies that gender differences in full CTC relief eligibility should be irrelevant. Pre-2018, the maximum envisioned per-child benefit ($1,000) was available as both a refund (ACTC) and a credit (CTC). Given that full relief realization, regardless of relief pathway (ACTC versus CTC), is an evolving goal of the CTC program, particularly for low-income households, our discussion thus far may overstate gender differences in full relief eligibility. To address this critique, we subsequently assess eligibility for the full magnitude of relief. Doing so demonstrates that single mothers were—given a lower refundability threshold—more likely to realize full program benefits relative to single fathers (Fig. 4a, 4b).

However, this finding is misleading absent consideration of which relief pathway is utilized. Although single mothers are overrepresented in full relief eligibility, this is driven by their overrepresentation in full ACTC eligibility (Fig. 5a, 5b). Isolating relief eligibility via non-refundable mechanisms reveals that single fathers are overrepresented in full CTC eligibility, albeit to a lesser extent than single mothers' overrepresentation in full ACTC eligibility. Accounting for this consideration highlights - owing to lower income - single mothers' disproportionate reliance on refundability to realize full program benefits. This finding is timely given programmatic changes to the CTC in 2018 that introduced disparity between the ACTC and CTC ($1,400 and $2,000, respectively).

What drives income differences between single mothers and single fathers? While our findings emphasize income-driven gender differences in relief eligibility, based on our data source we cannot discern why these differences exist. Lower income among single mothers may reflect reduced labor force participation rates, which would be consistent with women taking on more household and childcare duties than men. This would subsequently limit work availability and by consequence, income (40-44). However, income differences may also reflect gender discrimination, managerial discretion, clustering around lower-paid, less flexible occupations and behavioral variation in task execution (45-48). The CPS' design is insufficient to determine which explanation (or combination thereof) is applicable.

Nevertheless, the manifestation of income-driven gender differences warrants scrutiny given the CTC's envisioned purpose as an income stabilization tool for all working parents, regardless of marital status or gender (5, 6). Given this purpose, how can inequities in relief eligibility be addressed?

The simplest pathway would entail eliminating the CTC's income requirement, an outcome achieved by the 2021 ARP. Not only did the ARP increase the maximum available CTC, but it also eliminated the need for labor force participation, a historical precursor to the CTC into both CTC relief eligibility and realization of the program's most generous benefits. Consequently, the CTC in 2021 is analogous to a universal child allowance where relief eligibility and relief magnitude are unaffected by, among other things, refundability thresholds, refundability rates and phaseout thresholds.

However, the permanence of an expanded CTC remains unclear, owing in part to concerns over labor supply effects and program expansion costs (12, 13, 15, 49). The latter issue is particularly relevant as initial cost estimates of transforming the CTC into an allowance vary between $1.1 and $1.6 trillion over a 10-year budget window (12, 49), figures that have drawn political opposition owing to deficit spending concerns. How can our findings inform these efforts? Would changes to the CTC facilitate increases in overall program accessibility and generosity? If so, which ones?

For poverty alleviation efforts, reducing the refundability threshold is the surest means of increasing program accessibility (50). Our model estimates that eliminating income requirements for program participation would increase access for 0.49 (0.30), 0.83 (0.59), and 1.95 (0.98) percent of married parents, single fathers and single mothers respectively. In aggregate, based on 2018 data and applicable parameters 311,000 (176,000) households with children would benefit from eliminating the refundability threshold.

However, increasing program generosity has—under the earned income formula—the opposite effect, namely a widening of relief equity disparities. A more generous CTC program causes a downward shift in relief eligibility owing to the pairing of relief eligibility to income thresholds (Table 1a). As benefit magnitude increases, fewer parents, particularly single mothers, qualify for it, which forces previously-eligible parents into less generous benefits categories. For example, increasing the magnitude of the CTC decreases the number of parents who qualify for the full CTC while increasing the number of parents who qualify for full ACTC relief.

To what extent does raising the magnitude of the ACTC address this inequity? One of our key findings (Section 3.1) is that prior to 2018, gender differences in relief realization have been ameliorated owing to numerical parity between the CTC and the ACTC. The ACTC has lower income thresholds ($11,833 in 2018) to claim the full benefit ($1,400), compared to the CTC ($36,950 to claim $2,000). What would happen were the maximum monetary benefits of the CTC program made available to program participants as both a credit (CTC) and a refund (ACTC) in 2018?

In 2018, 76.72 (64.29) percent of single fathers qualified for the full CTC relief (worth $2,000) compared to 58.44 (43.37) percent of single mothers. Were the ACTC raised from $1,400 to $2,000 (a move which raises the requisite income threshold for the ACTC from $11,833 to $15,833), our model estimates that 95 (92.03) percent of single fathers would qualify for maximum relief (as either a credit or refund) compared to 87.15 (80.35) percent of single mothers. Consequently, single parent gender differences in realizing the maximum available monetary relief would decrease from 18.28 (20.92) percent to 7.85 (11.68) percent (Table 3c). Gender differences persist, albeit to a lesser extent, when the $2,000 benefit is realized as a combination of a credit and refund. Under these conditions, differences decrease from 12.71 (13.56) percent to 7.85 (11.68).

Consequently, our fourth finding is that gender differences in full relief realization can be mitigated via parity between the ACTC and CTC. Although ACTC magnitude increases cause a downward shift in relief eligibility (fewer single mothers now qualify for the full ACTC), a higher percentage of single mothers become eligible to claim the same magnitude of relief (via the CTC and the ACTC combined). Our model predicts that the result is a narrowing of the gap in maximum relief eligibility between single fathers and single mothers.

We acknowledge and caution however, that ACTC-CTC parity risks widening gender inequity if parity is paired with increasing the magnitude of relief (Fig. 6a,6b). Given that the CTC's design necessitates – under the earned income formula - higher income in exchange for more relief, a larger envisioned per-child benefit invariably prices out more parents as income requirements become more stringent. This is particularly detrimental to single mothers given their income profile. This outcome warrants consideration given policy proposals that increase the magnitude of CTC-related benefits while leaving unaltered other aspects of the CTC's design (51-53).

## 4.0 Limitations and Conclusion

Limitations to our approach warrant discussion.

First, while instating numerical parity between the CTC and ACTC represents a pathway toward achieving gender relief equity, doing so may impact labor supply (13, 15). Lowering income thresholds for full relief realization increases the marginal effective tax rate as labor becomes less valuable than leisure. This is particularly relevant given evidence that low-income workers—specifically, those targeted by the CTC program—have higher labor supply elasticities than other workers, especially in the component of their labor response that reflects movement in and out of the workforce (54). Given numerical parity between the CTC and ACTC, the prevalence of these effects and their consequences warrants further scrutiny.

Second, although our work examines numerical parity between the ACTC and CTC as a pathway towards rectifying inequities in relief realization, other alternatives admittedly do exist and also warrant scrutiny (55-57). These include lowering the refundability threshold and increasing the refundability rate, both of which work to increase program access and relief eligibility simultaneously (30,55-57). Future work should examine the viability of these alternatives relative to ACTC-CTC parity and quantify externalities—particularly cost considerations—associated with these alternatives. Quantifying labor supply differentials across pathways is also timely. We note however that changes to the refundability threshold and refundability rate would not, *ceteris paribus*, address documented inequities in full relief realization.

Third, the disproportionate realization of full CTC relief owing to marriage should not be construed as an endorsement of marriage, particularly among low-income parents. Studies show marriage among economically vulnerable couples may impose negative externalities of its own. These include greater fluctuations in spousal satisfaction, increased variability in satisfaction levels between spouses, and most notably, a higher likelihood of divorce (when compared to higher-income couples) (58). These findings warrant scrutiny given ambiguity over the extent to which the CTC program may facilitate meaningful reductions in poverty (13).

Finally, we note that our reliance on the CPS allows for estimating relief eligibility rather than the precise estimation of the relief magnitude realized. Further work should investigate this issue coupled with stratification of relief magnitude by categories such as race, education, and occupation. Furthermore, we reiterate that S1 and S2 respectively represent the most conservative and moderate scenarios produced by accounting for known error sources stemming from the CPS' design. Consequently, our estimates represent upper and lower bounds of the actual effect sizes. More accuracy warrants further study. However, we note that this pursuit would serve the interests of precision but would not impact our findings, as marital and gender effects presented thus far are consistent across scenarios.

Limitations notwithstanding, our results show clear, consistent, and compelling evidence of inequities in CTC relief realization. Owing to the CTC's income requirement, these inequities are concentrated among working parents, particularly single mothers, in lower segments of the income distribution. Given these findings, the intent of the CTC program warrants scrutiny. If that intent entails providing income stabilization to working parents at large, the CTC has (and continues to) achieve this intention (1). If, however, the CTC's intent program's goal is to aid economically vulnerable parents, further programmatic changes may be necessary (5, 6). We caution that under the CTC's income requirement, program accessibility and program generosity act as opposing forces. Increasing accessibility via a lower refundability threshold imposes a less onerous income requirement, whereas increasing generosity, when realized as a credit in particular, necessitates higher income.

To the extent that the CTC's envisioned intent is financial assistance for low-income parents, we emphasize that meeting these income requirements make relief realization more challenging for the very individuals it is increasingly envisioned to help.

**Acknowledgment**

This work was supported by the Robin Hood Foundation. The authors thank Jason Cone, Loris Toribio, Jonathan Bydlak, Nick Johnson, Art Lander, and Nimblebot LLC for their facilitation of this work. We are particularly indebted to Alec Kunzwiler and R.J. Caruso Tax and Accounting for income estimation assistance. The findings and conclusions expressed are solely those of the authors and do not represent the views of the funding or affiliated organizations.

**References**

1. "General Explanation of Tax Legislation Enacted in 1997*." Joint Committee on Taxation*, 17 Dec. 1997.

2. "The Child Tax Credit: Legislative History." *Congressional Research Service*, 23 December 2021. https://crsreports.congress.gov/product/pdf/R/R45124

3. Brooks-Gunn, Jeanne, and Greg J. Duncan. "The Effects of Poverty on Children." *The Future of Children*, vol. 7, no. 2, Princeton University, 1997, pp. 55–71, https://doi.org/10.2307/1602387.

4. Chaudry, Ajay, and Christopher Wimer. "Poverty Is Not Just an Indicator: The Relationship between Income, Poverty, and Child Well-Being." *Academic Pediatrics*, vol. 16, no. 3, ser. 23-29, 1 April 2016, pp. 23-29, https://doi.org/10.1016/j.acap.2015.12.010.

5. Parolin, Zachary, et al. "The Initial Effects of the Expanded Child Tax Credit on Material Hardship." *National Bureau of Economic Research Working Papers*, Sept. 2021, https://doi.org/10.3386/w29285.

6. National Academies of Sciences, Engineering, and Medicine. *A Roadmap to Reducing Child Poverty.* Washington, DC: The National Academies Press, 2019, https://doi.org/10.17226/25246.

7. Curran, M.A. "Catching Up on the Cost of Raising Children: Creating an American Child Allowance." *Big Ideas: Pioneering Change, Innovative Ideas for Children and Families.* Washington DC: First Focus, 2015.

8. Garfinkel, Irwin, et al. "Doing More for Our Children: Modeling a Universal Child Allowance or More Generous Child Tax Credit." *The Century Foundation*, 16 March 2016, https://tcf.org/content/report/doing-more-for-our-children/?session=1.

9. Shaefer, H Luke et al. "A Universal Child Allowance: A Plan to Reduce Poverty and Income Instability Among Children in the United States." *The Russell Sage Foundation journal of the social sciences: RSF* vol. 4, no.2, 2018, pp. 22-42, doi:10.7758/RSF.2018.4.2.0.

10. Sherman, Arloc, et al. "Earnings Requirement Would Undermine Child Tax Credit's Poverty-Reducing Impact While Doing Virtually Nothing to Boost Parents' Employment." *Center on Budget and Policy Priorities*, 23 Sept. 2021, https://www.cbpp.org/research/federal-tax/earnings-requirement-would-undermine-child-tax-credits-poverty-reducing-impact.

11. Harris, David Brooks. "The Child Tax Credit: How the United States Underinvests in Its Youngest Children in Cash Assistance and How Changes to the Child Tax Credit Could Help." *Columbia University*, 2012.

12. Garfinkel, Irwin, et al. "The Costs and Benefits of a Child Allowance." *Center on Poverty and Social Policy,* v. 5, n. 1, Columbia University, 18 Feb. 2021, https://www. povertycenter.columbia.edu/news-internal/2021/child-allowance/cost-benefit-analysis.

13. Corinth, Kevin et al. "The Anti-Poverty, Targeting, and Labor Supply Effects of the Proposed Child Tax Credit Expansion" *Becker Friedman Institute,* no. 2021-115, University of Chicago Economics, 6 October 2021, https://doi.org/10.3386/w29366.

14. Congressional Research Service. *"*The Child Tax Credit: How it Works and Who Receives It." *Congressional Research Service,* 12 Jan. 2021.

15. Winship, Scott. "The conservative case against child allowances." *American Enterprise Institute*, 5 March 2021, https://www.aei.org/research-products/report/the-conservative-case-against-child-allowances/.

16. Schirle, Tammy. "The effect of universal child benefits on labour supply." *Canadian Journal of Economics/Revue canadienne d'économique,* vol.48, no.2, 2015, pp. 437-463.

17. Marinescu, Ioana. "No strings attached: The behavioral effects of US unconditional cash transfer programs." *National Bureau of Economic Research*, 2018.

18. Robins, Philip K. "A Comparison of the Labor Supply Findings from the Four Negative Income Tax Experiments." *The Journal of Human Resources*, vol. 20, no. 4, 1985, pp. 567–582., https://doi.org/10.2307/145685.

19. Moffitt, Robert A. "The Negative Income Tax: Would It Discourage Work?" *Monthly Labor Review*, Apr. 1981, pp. 23–27.

20. Widerquist, Karl. "A Failure to Communicate: What (If Anything) Can We Learn from the Negative Income Tax Experiments?" *The Journal of Socio-Economics*, vol. 34, no. 1, Feb. 2005, pp. 49–81., https://doi.org/10.1016/j.socec.2004.09.050.

21. Eissa, Nada, and Jeffrey B. Liebman. "Labor supply response to the earned income tax credit." *The Quarterly Journal of Economics,* vol.111, no.2, 1996, pp. 605-637.

22. Baker, Michael et al. "The Effects of Child Tax Benefits on Poverty and Labor Supply: Evidence from the Canada Child Benefit and Universal Child Care Benefit." *National Bureau of Economic Research*, 2021.

23. Jones, Damon, and Ioana Marinescu. "The labor market impacts of universal and permanent cash transfers: Evidence from the Alaska Permanent Fund." *National Bureau of Economic Research*, 2018.

24. Kidd, Andrew. "The Effect of the Child Tax Credit on the Labor Supply of Mothers." *University of Notre Dame*, last accessed Feb. 2 2022, https://economics.nd.edu/assets/41472.

25. Zheng, Wei. "Child Tax Credit and Maternal Labor Supply." University of Connecticut, Sept. 2020.

26. Hammond, Samuel, and Robert Orr. "Report: Measuring the Child Tax Credit's Economic and Community Impact." *Niskanen Center*, Niskanen Center, 19 Oct. 2021, https://www.niskanencenter.org/report-measuring-the-child-tax-credits-economic-and-community-impact/.

27. Hartley, Robert Paul, et al. "The Benefits and Costs of a U.S. Child Allowance." *NBER*, National Bureau of Economic Research, 21 Mar. 2022, http://www.nber.org/papers/w29854.

28. Garfinkel, Irwin, et al. "The Benefits and Costs of a U.S. Child Allowance." *NBER*, National Bureau of Economics Research, 21 Mar. 2022, http://www.nber.org/papers/w29854

29. Goldin, Jacob, and Katherine Michelmore. "Who Benefits from the Child Tax Credit?" *NBER Working Paper Series*, v. 27940, Oct. 2020, https://doi.org/10.3386/w27940.

30. Greenstein, Robert, et al. "Improving the Child Tax Credit for Very Low-Income Families." *The US Partnership on Mobility from Poverty*, Urban Institute, Apr. 2018, https://www.mobilitypartnership.org/improving-child-tax-credit-very-low-income-families.

31. Collyer, Sophie, et al. Vol. 3, Columbia Population Research Center, 2019, *Left Behind: The One-Third of Children in Families Who Earn Too Little to Get the Full Child Tax Credit.*

32. Curran, Megan, and Sophie Collyer. Vol. 4, Columbia Population Research Center, 2020, Children Left Behind in Larger Families: The Uneven Receipt of the Federal Child Tax Credit by Children's Family Size.

33. "Current Population Survey Design and Methodology Technical Paper 77." *U.S. Census Bureau*, 2019. https://www2.census.gov/programs-surveys/cps/methodology/CPS-Tech-Paper-77.pdf.

34. "Current Population Survey, Annual Social and Economic Supplement." *US Census Bureau*, 2018. https://www.census.gov/data/datasets/time-series/demo/cps/cps-asec.2018.html

35. "Annual Social and Economic Supplement (ASEC) of the Current Population Survey (CPS)." *US Census Bureau*, 8 Oct. 2021, https://www.census.gov/programs-surveys/saipe/guidance/model-input-data/cpsasec.html.

36. Moore, Jeffrey C., et al. "Income Measurement Error in Surveys: A Review." *United States Census Bureau*, 1997, https://www.census.gov/content/dam/Census/library/working-papers/1997/adrm/sm97-05.pdf

37. Ploeg, Michele Ver, et al. *Studies of Welfare Populations: Data Collection and Research Issues*. National Academy Press, 2002.

38. "Poverty Thresholds by Size of Family and Number of Children." *US Census Bureau*, 8 Oct. 2021, https://www.census.gov/data/tables/time-series/demo/income-poverty/historical-poverty-thresholds.html.

39. "Subject Definitions." *US Census Bureau*, 8 Oct. 2021, https://www.census.gov/programs-surveys/cps/technical-documentation/subject-definitions.html#:~:text=Own%20children%20in%20a%20family,or%20parent%20in%20the%20subfamily.

40. Bolotnyy, Valentin and Natalia Emanuel, "Why Do Women Earn Less Than Men? Evidence from Bus and Train Operators." *Harvard Faculty of Arts and Science*, March 2021, https://scholar.harvard.edu/files/bolotnyy/files/be_gender_gap.pdf.

41. Parker, Kim, et al. "Raising Kids and Running a Household: How Working Parents share the Load." *Pew Research Trust*, November 2015, https://www.pewresearch.org/social-trends/wp-content/uploads/sites/3/2015/11/2015-11-04_working-parents_FINAL.pdf.

42. Bertrand, Marianne, et al. "Gender Identity and Relative Income Within Households." *National Bureau of Economic Research*, May 2013, https://www.nber.org/system/files/working_papers/w19023/w19023.pdf.

43. Noonan, Mary C, et al. "Pay Differences Among the Highly Trained: Cohort Differences in the Sex Gap in Lawyers' Earnings." *Social Forces,* vol. 84, no. 2, 2005, pp. 853–872, https://myweb.uiowa.edu/noona/sf.pdf.

44. Reyes, Jessica Wolpaw. "Reaching Equilibrium in the Market for Obstetricians and Gynecologists." *American Economic Review*, vol. 97 no. 2, 2007, pp. 407–411.

45. Blau, Francine D. and Lawrence M. Kahn. "The Gender Wage Gap: Extent, Trends, and Explanations." *Journal of Economic Literature*, vol. 55 no. 3, 2017, pp. 789–865.

46. Lazear, Edward P and Sherwin Rosen. "Male-Female Wage Differentials in Job Ladders." *Journal of Labor Economics*, vol. 8 no. 2, 1990, pp. S106–S123.

47. Cohen, Philip N. and Matt L Huffman. "Working for the Woman? Female Managers and the Gender Wage Gap." *American Sociological Review*, vol. 72 no. 5, 2007, pp. 681–704.

48. Cook, Cody, et al. "The Gender Earnings Gap in the Gig Economy: Evidence from over a Million Rideshare Drivers." *NBER Working Paper Series*, vol. 24732, June 2018, http://www.nber.org/papers/w24732.

49. York, Erica, and Huaqun Li. "Making the Expanded Child Tax Credit Permanent Would Cost Nearly $1.6 Trillion." *Tax Foundation*, 30 Aug. 2021, https://taxfoundation.org/expanded-child-tax-credit-permanent/.

50. Ortigueira, Salvador, and Nawid Siassi. "Income Assistance, Marriage, and Child Poverty: An Assessment of the Family Security Act." *Economic Modelling*, vol. 111, 14 Mar. 2022, p. 105827., https://doi.org/10.1016/j.econmod.2022.105827.

51. Maag, Elaine, and Ramirez, Elena. "Reforming the Child Tax Credit: An Update." *Tax Policy Center*, Urban Institute and Brookings Institution, 19 Oct. 2016, https://www.taxpolicycenter.org/publications/reforming-child-tax-credit-update/full

52. Maag, Elaine, and Austin, Lydia. "Implications for Changing the Child Tax Credit Refundability Threshold." *Tax Policy Center*, Urban Institute and Brookings Institution, 24 July 2014, https://www.taxpolicycenter.org/publications/implications-changing-child-tax-credit-refundability-threshold.

53. Maag, Elaine. "Issues in Child Benefit Administration in the United States." *Urban Institute*, Urban Institute, Dec. 2021, https://www.urban.org/sites/default/files/publication/105218/issues-in-child-benefit-administration-in-the-united-states_1.pdf.

54. McClellend, Robert, and Shannon Mok. "A Review of Recent Research on Labor Supply Elasticities." *Congressional Budget Office*, Oct. 2012, https://www.cbo.gov/sites/default/files/112th-congress-2011-2012/workingpaper/10-25-2012-recentresearchonlaborsupplyelasticities.pdf.

55. Thomas, Adam, and Isabel V. Sawhill. "A Tax Proposal for Working Families with Children." *Brookings*, Brookings, 28 July 2016, https://www.brookings.edu/research/a-tax-proposal-for-working-families-with-children/.

56. Cass, Oren, and Wells King. "The Family Income Supplemental Credit Expanding the Social Compact for Working Families." *American Compass*, Feb. 2021, https://americancompass.org/wp-content/uploads/2021/02/American_Compass-The_Family_Income_Supplemental_Credit-2021Feb18.pdf.

57. Winship, Scott. American Enterprise Institute, 2021, *Reforming Tax Credits to Promote Child Opportunity and Aid Working Families*, https://www.aei.org/research-products/report/reforming-tax-credits-to-promote-child-opportunity-and-aid-working-families/.

58.  Jackson, Grace L., et al. "Household Income and Trajectories of Marital Satisfaction in Early Marriage*." Journal of Marriage and Family*, vol. 79, no. 3, Jan. 31 2017, pp. 690–704., doi:10.1111/jomf.12394.

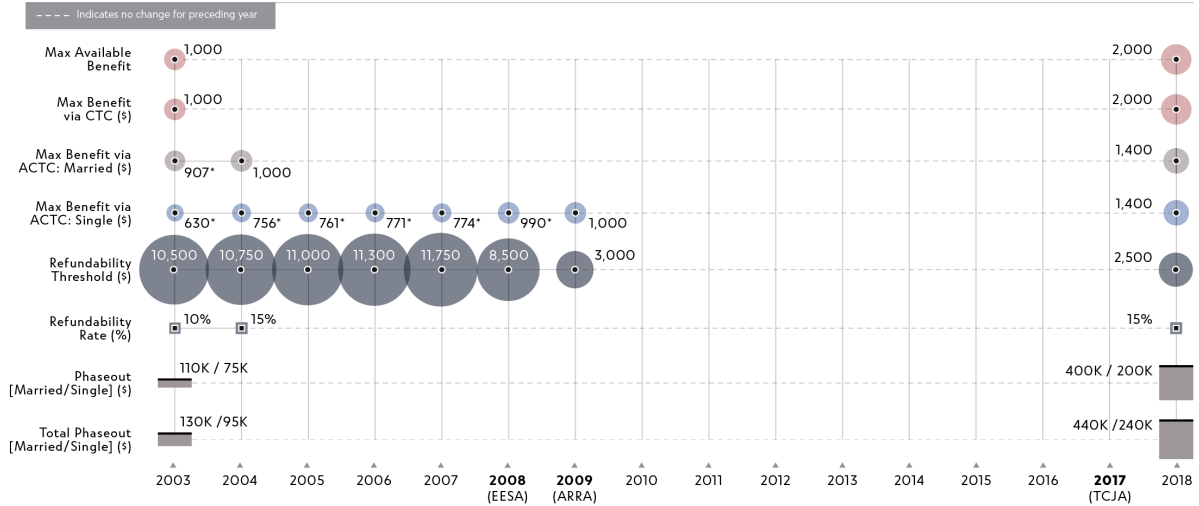

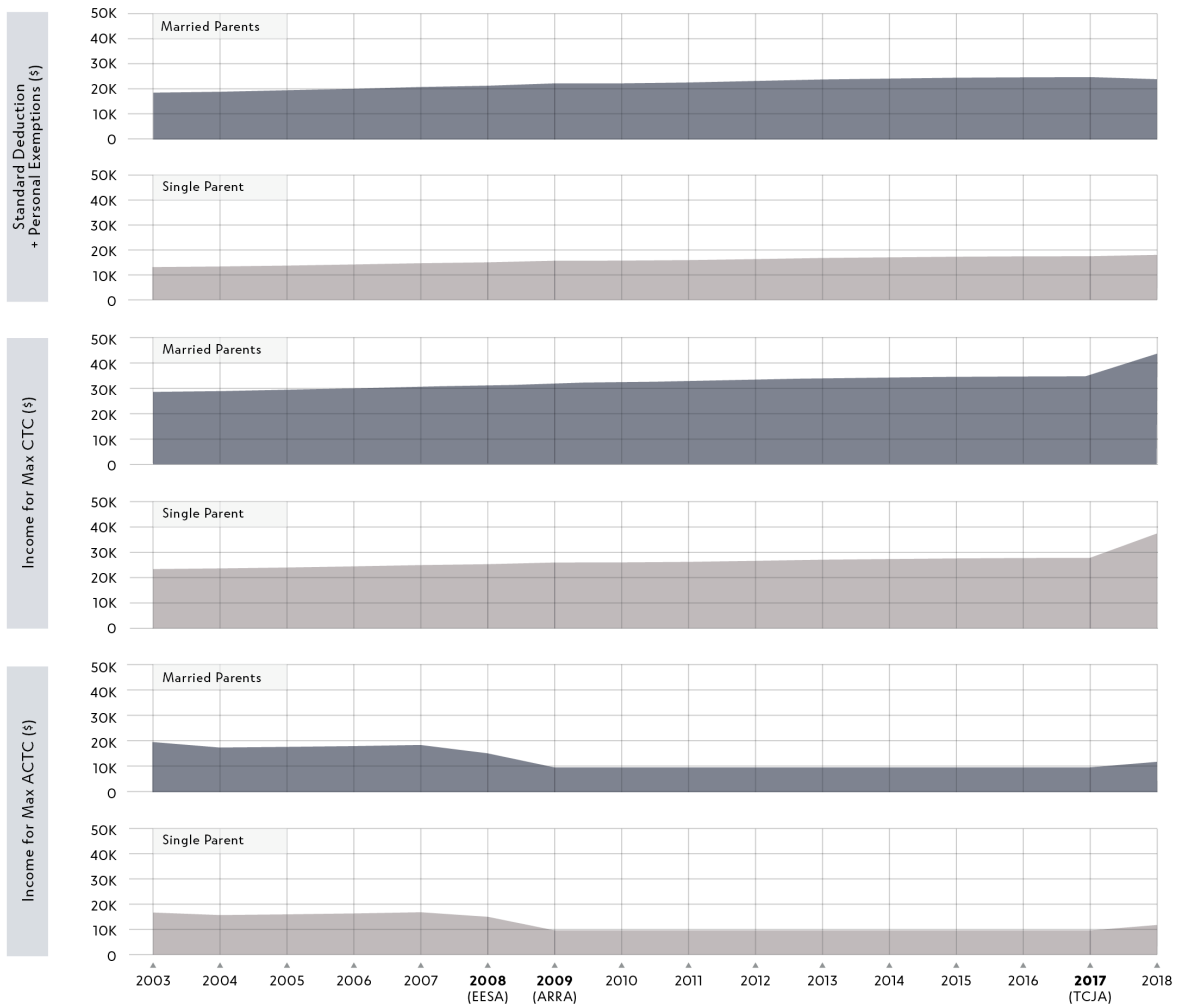

**Figure 1: CTC program parameters and estimated income thresholds (S1)**

| | | Married Parents | Single Fathers | Single Mothers |
|---|---|---|---|---|
| **Step 1** | *2018 distribution by 2018 program standards* | 74.37 [58.26] | 76.72 [64.29] | 58.44 [43.37] |
| **Step 2** | *2018 distribution by 2017 program standards* | 84.75 [76.67] | 60.52 [52.30] | 57.75 [47.71] |
| **Step 3** | *2018 distribution by 2017 program standards + changing CTC from $1,000 to $2,000* | 74.37 [58.26] | 50.05 [37.28] | 42.56 [30.92] |
| **Step 4** | *2018 distribution by 2017 program standards + changing CTC from $1,000 to $2,000 + standard deduction adjustment* | 74.37 [61.86] | 50.05 [37.28] | 42.56 [30.92] |
| **Step 5** | *2018 distribution by 2017 program standards + changing CTC from $1,000 to $2,000 + Standard deduction adjustment + refundability threshold change to 2018 standard* | 74.37 [61.86] | 50.05 [37.28] | 42.56 [30.92] |
| **Step 6** | *2018 distribution by 2017 program standards + changing CTC from $1,000 to $2,000 + Standard deduction adjustment + refundability threshold change to 2018 standard + phaseout change to 2018 standard* | 74.37 [61.86] | 76.72 [67.92] | 58.44 [49.05] |
| **Step 7** | *2018 distribution by 2017 program standards + changing CTC from $1,000 to $2,000 + Standard deduction adjustment + refundability threshold change to 2018 standard + phaseout change to 2018 standard + ACTC change from $1,000 to $1,400* | 74.37 [61.86] | 76.72 [67.92] | 58.44 [49.05] |
| **Step 8** | *2018 distribution by 2017 program standards + changing CTC from $1,000 to $2,000 + Standard deduction adjustment + refundability threshold change to 2018 standard + phaseout change to 2018 standard + ACTC change from $1,000 to $1,400 + TCJA tax rate adjustments + adjustment for year-on-year dependent heterogeneity* | 74.37 [58.26] | 76.72 [64.29] | 58.44 [43.37] |

**Table 1a: Full CTC Relief Eligibility Impact Analysis for 2018**

|  |  | Married Parents | Single Fathers | Single Mothers |
|---|---|---|---|---|
| **Step 1** | *2018 distribution by 2018 program standards* | 23.76 [37.49] | 21.03 [30.71] | 34.05 [43.78] |
| **Step 2** | *2018 distribution by 2017 program standards* | 14.08 [20.85] | 11.49 [14.02] | 21.61 [24.55] |
| **Step 3** | *2018 distribution by 2017 program standards + changing ACTC from $1,000 to $1,400* | 13.37 [19.09] | 10.57 [12.06] | 18.86 [18.73] |
| **Step 4** | *2018 distribution by 2017 program standards + changing ACTC from $1,000 to $1,400 + standard deduction change to 2018 standard* | 13.37 [19.09] | 11.90 [12.06] | 21.69 [21.72] |
| **Step 5** | *2018 distribution by 2017 program standards + changing ACTC from $1,000 to $1,400 + standard deduction change to 2018 standard + changing refundability threshold from $3,000 to $2,500* | 13.37 [19.09] | 11.90 [12.06] | 21.69 [24.31] |
| **Step 6** | *2018 distribution by 2017 program standards + changing ACTC from $1,000 to $1,400 + standard deduction change to 2018 standard + changing refundability threshold from $3,000 to $2,500 + changing phaseout thresholds to 2018* | 13.37 [19.09] | 11.90 [12.06] | 21.69 [24.31] |
| **Step 7** | *2018 distribution by 2017 program standards + changing ACTC from $1,000 to $1,400 + standard deduction change to 2018 standard + changing refundability threshold from $3,000 to $2,500 + changing phaseout thresholds to 2018 + changing CTC from $1,000 to $2,000* | 23.76 [33.89] | 21.03 [27.08] | 34.05 [38.11] |
| **Step 8** | *2018 distribution by 2017 program standards + changing ACTC from $1,000 to $1,400 + standard deduction change to 2018 standard + changing refundability threshold from $3,000 to $2,500 + changing phaseout thresholds to 2018 + changing CTC from $1,000 to $2,000 + TCJA tax rate adjustments + adjustment for year-on-year dependent heterogeneity* | 23.76 [37.49] | 21.03 [30.71] | 34.05 [43.78] |

**Table 1b: Full ACTC Relief Eligibility Impact Analysis for 2018**

|  |  | Married Parents | Single Fathers | Single Mothers |
|---|---|---|---|---|
| **Step 1** | *2018 distribution by 2018 program standards* | 74.37 [58.26] | 76.72 [64.29] | 58.44 [43.37] |
| **Step 2** | *2018 distribution by 2018 program standards + changing ACTC from $1,400 to $2,000* | 96.48 [92.74] | 95 [92.03] | 87.15 [80.35] |
| **Step 3** | *2018 distribution by 2018 program standards + changing ACTC from $1,400 to $2,000 + eliminating refundability threshold* | 97.51 [92.74] | 96.96 [94.30] | 90.39 [84.57] |

**Table 1c: Percentage of Each Parental Group qualifying for Full Relief (as either a credit or refund)**

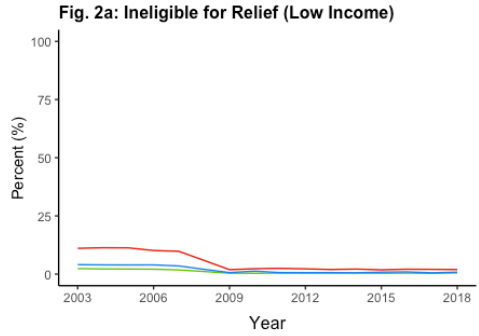

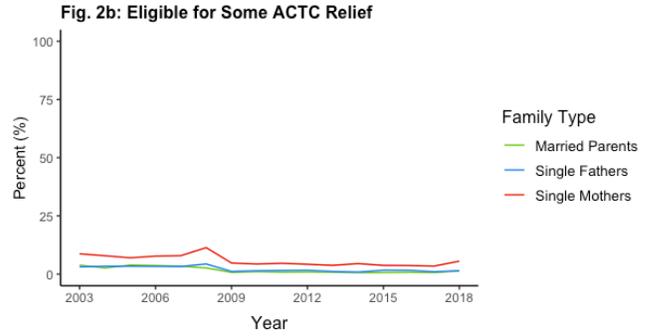

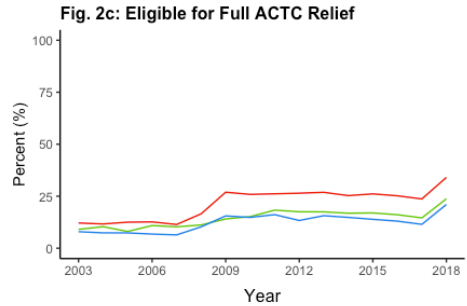

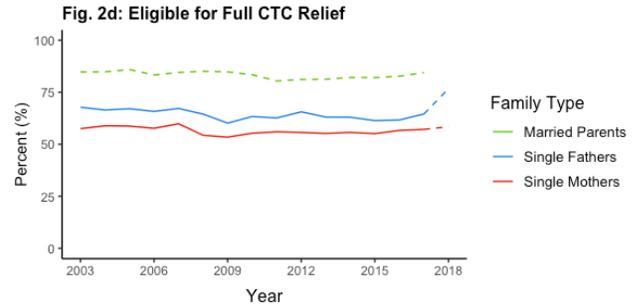

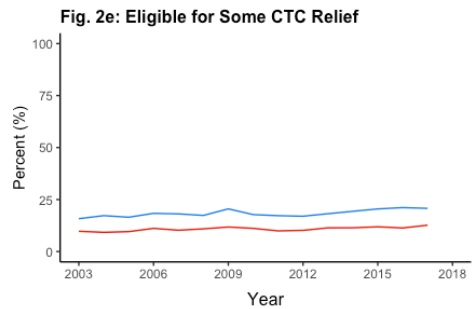

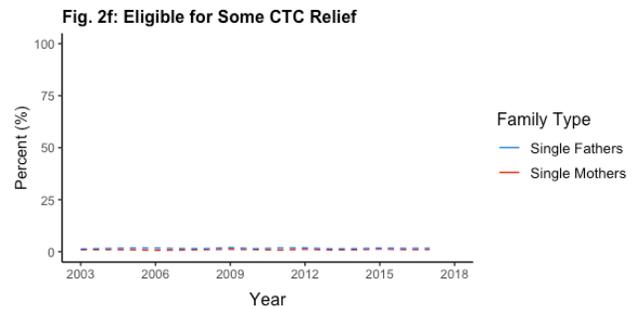

**Figure 2: Relief eligibility estimates by parental group (S1). Solid lines represent values generalizable to the population. Dotted lines indicate underestimated population-wide values. Values that cannot be estimated are omitted from presentation/analysis.**

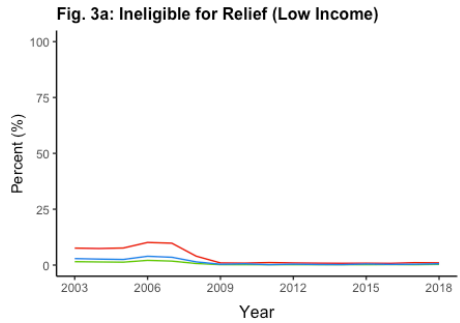

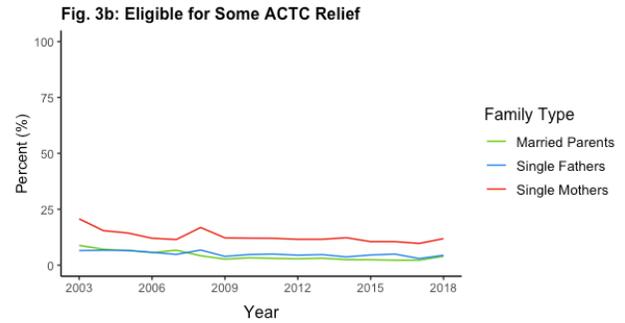

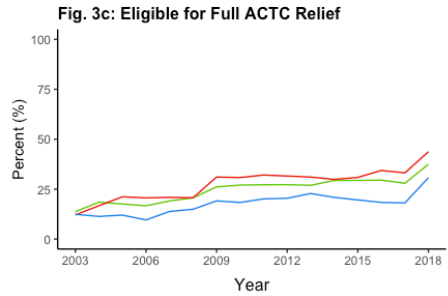

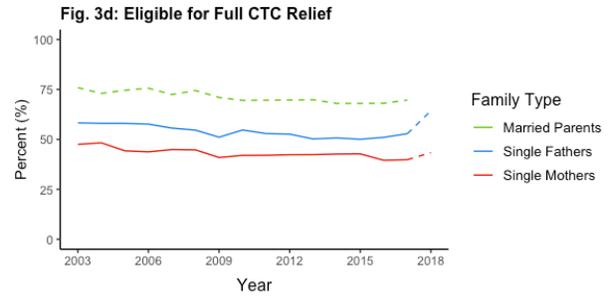

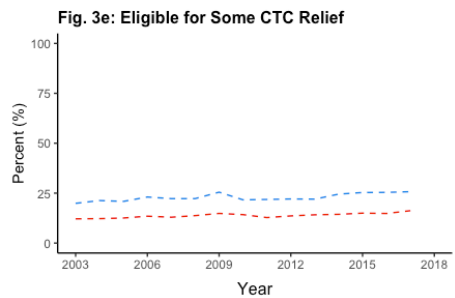

**Figure 3: Relief eligibility estimates by parental group (S2). Solid lines represent values generalizable to the population. Dotted lines indicate underestimated population-wide values. Values that cannot be estimated are omitted from presentation/analysis.**

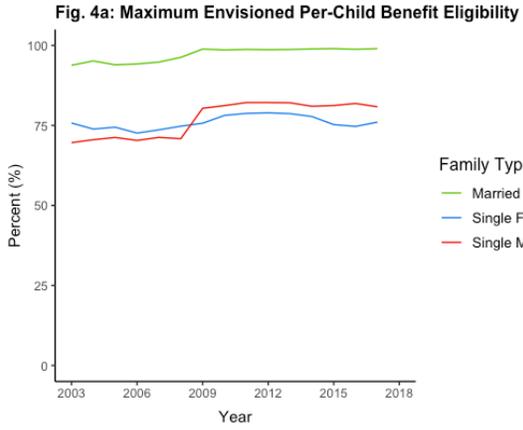
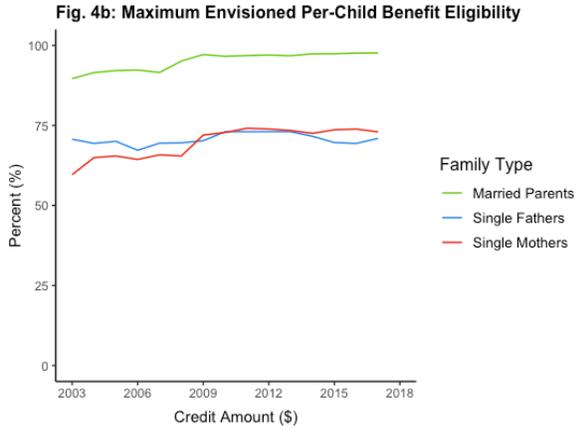

**Figure 4: Relief eligibility for maximum envisioned per-child benefit. A and B correspond to S1 and S2 scenario respectively.**

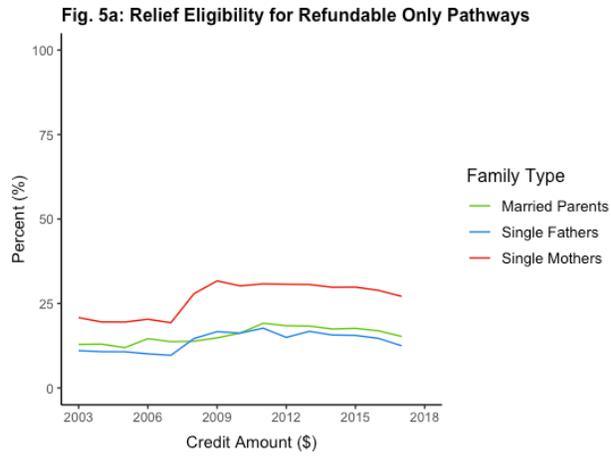
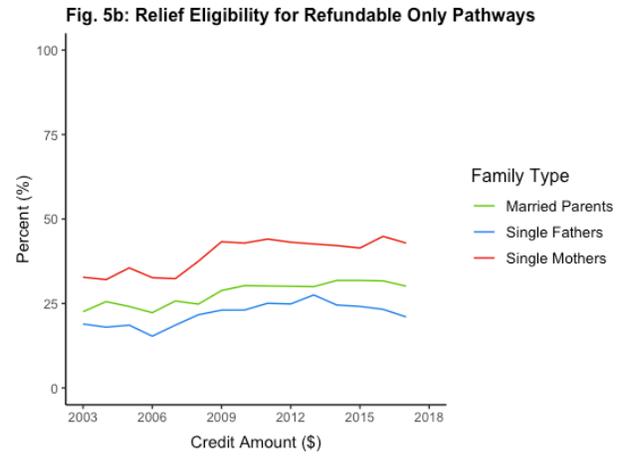

**Figure 5: Relief eligibility for refundable only pathways. A and B correspond to S1 and S2 scenario respectively.**

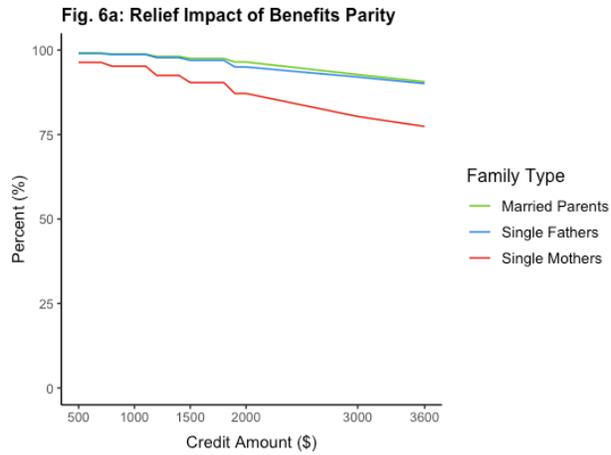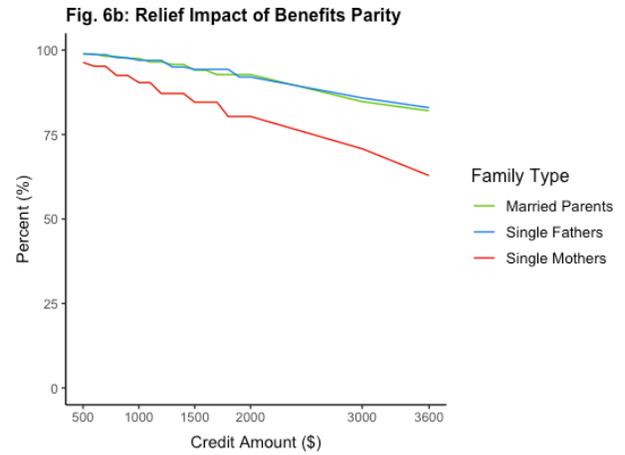

**Figure 6: Impact of ACTC-CTC parity by credit size. A and B correspond to S1 and S2 scenario respectively.**

**Supplementary Information**

In this document, we provide a detailed overview of our analysis and underlying parameters. In Section I, we construct our datasets and chosen scenarios and justify our assumptions. In Section II, we detail our income distribution analysis. In Section III, we describe our statistical and comparative analyses for 2003 - 2017 and subsequently, 2017 - 2018.

**Section I**

In this section, we discuss our approach to estimating the total number of married parents, single mothers, and single fathers who qualify for child tax credit (CTC) relief. This entails analysis of data from the U.S. Census Bureau Current Population Survey (CPS), which is a nationally-representative database sampling more than 75,000 households. The CPS' Annual Social and Economic Supplement (ASEC) surveys households on demographic characteristics including the respondent's income, marital status, and number of "related children" (defined as individuals under 18 who are share either a biological or legal relationship with the respondent e.g., birth, adoption, marriage) (1-3).

We levy the CPS Table Creator to create data tables of ASEC responses for each year from 2003 to 2018. In particular, we focus on the number of Adult Civilian Workers (defined as workers aged 18+ who currently reside in the US and are not institutionalized) who have at least one child under 18 in the household. We emphasize this population as a proxy for the total number of parents in the workforce (4).

To estimate the total number of married parents, single mothers, and single fathers, we group individuals within this population according to their marital status (single or married) and sex (male or female), as well as their annual household gross income (measured from $0 to $100,000 in $2,500 increments). We subsequently combine married males and married females into a single category; namely, "married parents."

Additionally, we leverage the CPS to estimate each parental group's average number of dependents, defined as children below the age of 18. We do so to account for the varying size of tax deductions, as such tax-free relief was offered to all taxpayers as a function of their quantity of dependents. These deductions were subtracted from individuals' adjusted gross income, which subsequently impacted income thresholds required to realize full CTC benefits. Consequently, our model accounts for this parameter.

Owing to specific language used by the CPS, to account for potential overestimates of reported children per household we estimate an upper and middle bound for the number of children. Our upper-bound estimate presumes each household has at least one dependent, whereas our middle-bound estimates leverage responses to earlier CPS questions to calculate the total number of related dependents for each parental group.

Specifically, each respondent can indicate a discrete number between 0 and 7, or 8 or more, for the number of related children under 18 in the family (1). The CPS Table Creator also provides data on parents' relationship type and income bracket. We calculate the total number of children in each relationship group (i.e., single mother, single father, married parents) by multiplying the number of respondents by the number of children indicated and summing the products. Responses of "8 or more" children are treated as "8" in our calculations.

We subsequently calculate the average number of children in each parental group by dividing the total number of children in that group by the number of parents. We repeat this process year-on-year and estimate an average across all years. Table 1 shows our estimated number of dependents for each parental group.

| | Married Parents | Single Fathers | Single Mothers |
|---|---|---|---|
| **2003** | 1.89 | 1.68 | 1.72 |
| **2004** | 1.90 | 1.71 | 1.75 |
| **2005** | 1.90 | 1.71 | 1.76 |
| **2006** | 1.89 | 1.68 | 1.74 |
| **2007** | 1.89 | 1.68 | 1.73 |
| **2008** | 1.90 | 1.69 | 1.75 |
| **2009** | 1.89 | 1.67 | 1.74 |
| **2010** | 1.89 | 1.68 | 1.74 |
| **2011** | 1.88 | 1.67 | 1.71 |
| **2012** | 1.89 | 1.67 | 1.73 |
| **2013** | 1.89 | 1.69 | 1.74 |
| **2014** | 1.88 | 1.71 | 1.73 |
| **2015** | 1.90 | 1.69 | 1.73 |
| **2016** | 1.89 | 1.70 | 1.74 |
| **2017** | 1.89 | 1.66 | 1.74 |
| **2018** | 1.89 | 1.69 | 1.75 |
| **Average** | **1.89** | **1.69** | **1.74** |

Note: The left-most column denotes the corresponding year of each estimate; the top row displays the corresponding parental group.

**Table 1: Average Number of Children (S2 Only)**

After estimating counts of each parental group and their associated number of children, we then use tax guidelines from the Internal Revenue Service to calculate each group's requisite income threshold to qualify for CTC relief (5). For each group, year-on-year we calculate requisite income thresholds that correspond to the following categories of CTC relief: (a) ineligible for any relief (owing to low income), (b) eligible for some ACTC relief; (c) eligible for full ACTC relief; (d) eligible for full CTC relief; (e) eligible for some CTC relief; and (f) ineligible for any relief (owing to high income).

To provide robust estimates of parental groups' relief eligibility, we consider two additional factors: namely, tax filing status and CTC program parameters. We assume married parents file a single joint tax return, whereas single parents file as heads of household (6). Based on government sources, we also consider the impact of CTC program parameters such as ACTC and CTC magnitude, refundability rates, and refundability thresholds on requisite income thresholds (7).

Given the annual household incomes, average number of dependents, and requisite income thresholds for each parental group, we now approximate the number of married parents, single mothers, and single fathers who qualify for each of the previously discussed CTC relief categories. Analyzing the aforementioned parameters to assign parents to a CTC relief category, we use a single-parent household in 2009 as an illustrative example.

During this year, the refundability threshold was $3000, the refundability rate was 10 percent, and both the max ACTC and CTC were $1,000. In 2009, single parents with one child required $9,667 in income to realize the full ACTC and $25,650 for full CTC. Consequently, parents who earned between $9,667 and $25,650 are assigned to the full ACTC relief group, parents who earned less than $9,667 but more than $3,000 (the refundability threshold in 2009) are assigned to the some ACTC group, and those earning less than $3,000, to the no relief group (low income). Our model assigns single parents who earned more than $25,650 but less than $75,000 (the phaseout threshold for single parents in 2009) to the full CTC relief group. Single parents earning more than $75,000 remain eligible for some CTC relief (Category e) until reaching the total phaseout threshold (i.e., $95,000 for single parents in 2009). Parents who exceed the total phaseout threshold become ineligible for any relief owing to high income (Category f).

We repeat this process for each year (2003 – 2018) and parental group (married parents, single mothers, and single fathers) to estimate the number of parents who qualify for each eligibility category. However, limitations of the CPS and CPS Table Creator lead to potential errors in income estimation and income limit exclusion and thus require additional corrective measures.

First, CPS respondents indicate their income bracket, not their specific income. This results in income brackets occasionally overlapping with requisite income thresholds for CTC relief categories (e.g., respondents earning between $25,000 and $27,499 may or may not have satisfied the $25,650 threshold for full CTC relief in 2009). Given the lack of granularity in the CPS' income brackets, we are unable to determine to which relief category such respondents belonged. To account for this ambiguity, we calculate upper- and middle-bound estimates of parents eligible for relief.

We begin by constructing our upper-bound estimates. To do so, we presume households belonging to income bins that contain a requisite income threshold fail to meet said threshold. For example, if the requisite income threshold to claim full CTC relief is $25,650, we treat all households whose income bracket is $25,000-$27,499 as though their exact income was below $25,650. Consequently, such households are considered ineligible for full CTC relief and instead are included in the full ACTC category (i.e., Category c). Because this approach effectively raises each relief category's requisite income threshold, our upper-bound estimates are therefore conservative estimates of parental eligibility.

We then construct our middle-bound estimates. Here, households are assigned to relief categories based on whether the median of their income bracket exceeds the overlapping requisite income threshold. For example, households whose income bracket is $25,000-$27,499 would be treated as eligible for full CTC relief if the requisite income threshold were $25,650. However, if the requisite income threshold for full CTC were instead $27,000, these households would now be considered ineligible for full CTC relief and instead categorized as claiming full ACTC, as the requisite threshold now exceeds the income bracket's median (i.e., $26,250). Given that this approach does not inflate requisite income thresholds, our middle-bound estimates thus represent a more moderate estimate of parental relief eligibility.

To account for existing potential error sources (i.e., the average number of dependents and total number of parents qualifying for each category of CTC relief), we construct two scenarios. Our first scenario (hereafter, "S1") presumes all parents claim only one child as a dependent and uses our upper-bound estimates of relief eligibility. S1 thus represents the most conservative scenario possible by accounting for the aforementioned error sources.

Our second scenario (hereafter, "S2") draws upon our middle-bound estimates for the number of dependents (Table 1), year-on-year, for each parental group. Based on these figures, we recalculate each group's standard deduction and requisite income thresholds for all levels of CTC relief. We also leverage our middle-bound estimates of relief eligibility. S2 thus represents the most moderate scenario possible by accounting for the aforementioned error sources.

We intentionally construct the most conservative and moderate scenarios possible to account for any concerns regarding the accuracy of CPS responses. Owing to the methodology used to construct S1 and S2, effects are observed in S1 should be amplified in S2. By assessing various trends in both S1 and S2, we are thus able to ensure that our results are robust against artifacts produced by the CPS' design.

For each of S1 and S2, we subsequently estimate (using the aforementioned approach) the number of single mothers, single fathers, and married parents who qualify for each category of CTC relief. We then transform single fathers' counts into proportions using a two-step process. To begin, we sum the number of single fathers across all eligibility categories to calculate the total number of single fathers in a given year. We subsequently divide each count by this total to estimate, again for a given year, what proportion of single fathers qualified for each level of CTC and/or ACTC relief. We then repeat this process year-on-year from 2003 – 2018 for both single mothers and married parents. Our results are shown in Tables 2.1 and 2.2.

|   |    | 2018 | 2017 | 2016 | 2015 | 2014 | 2013 | 2012 | 2011 | 2010 | 2009 | 2008 | 2007 | 2006 | 2005 | 2004 | 2003 |
|---|----|------|------|------|------|------|------|------|------|------|------|------|------|------|------|------|------|
| **a** | MP | 0.0049 | 0.0033 | 0.0040 | 0.0033 | 0.0046 | 0.0042 | 0.0043 | 0.0042 | 0.0035 | 0.0038 | 0.0110 | 0.0178 | 0.0210 | 0.0218 | 0.0222 | 0.0235 |
|   | SF | 0.0083 | 0.0053 | 0.0093 | 0.0082 | 0.0054 | 0.0062 | 0.0059 | 0.0063 | 0.0126 | 0.0064 | 0.0204 | 0.0350 | 0.0394 | 0.0393 | 0.0395 | 0.0406 |
|   | SM | 0.0195 | 0.0203 | 0.0209 | 0.0183 | 0.0215 | 0.0196 | 0.0225 | 0.0246 | 0.0230 | 0.0190 | 0.0585 | 0.0978 | 0.1013 | 0.1126 | 0.1132 | 0.1106 |
| **b** | MP | 0.0139 | 0.0067 | 0.0080 | 0.0068 | 0.0062 | 0.0084 | 0.0089 | 0.0083 | 0.0106 | 0.0074 | 0.0263 | 0.0341 | 0.0368 | 0.0387 | 0.0262 | 0.0383 |
|   | SF | 0.0142 | 0.0101 | 0.0163 | 0.0167 | 0.0086 | 0.0112 | 0.0163 | 0.0155 | 0.0142 | 0.0113 | 0.0435 | 0.0326 | 0.0329 | 0.0335 | 0.0336 | 0.0309 |
|   | SM | 0.0557 | 0.0342 | 0.0367 | 0.0373 | 0.0451 | 0.0374 | 0.0422 | 0.0464 | 0.0433 | 0.0476 | 0.1137 | 0.0789 | 0.0769 | 0.0698 | 0.0787 | 0.0869 |
| **c** | MP | 0.2376 | 0.1457 | 0.1610 | 0.1695 | 0.1682 | 0.1748 | 0.1752 | 0.1831 | 0.1518 | 0.1407 | 0.1119 | 0.1027 | 0.1090 | 0.0804 | 0.1035 | 0.0905 |
|   | SF | 0.2103 | 0.1148 | 0.1302 | 0.1388 | 0.1481 | 0.1563 | 0.1333 | 0.1614 | 0.1481 | 0.1553 | 0.1027 | 0.0640 | 0.0680 | 0.0736 | 0.0738 | 0.0794 |
|   | SM | 0.3405 | 0.2368 | 0.2521 | 0.2612 | 0.2529 | 0.2687 | 0.2646 | 0.2616 | 0.2589 | 0.2692 | 0.1653 | 0.1140 | 0.1265 | 0.1252 | 0.1166 | 0.1213 |
| **d** | MP | 0.7437 | 0.8443 | 0.8271 | 0.8204 | 0.8210 | 0.8126 | 0.8115 | 0.8044 | 0.8342 | 0.8480 | 0.8508 | 0.8453 | 0.8331 | 0.8591 | 0.8481 | 0.8477 |
|   | SF | 0.7672 | 0.6456 | 0.6167 | 0.6138 | 0.6299 | 0.6306 | 0.6561 | 0.6263 | 0.6334 | 0.6017 | 0.6449 | 0.6723 | 0.6581 | 0.6709 | 0.6649 | 0.6781 |
|   | SM | 0.5844 | 0.5715 | 0.5666 | 0.5511 | 0.5571 | 0.5521 | 0.5569 | 0.5600 | 0.5532 | 0.5344 | 0.5432 | 0.5987 | 0.5767 | 0.5873 | 0.5889 | 0.5750 |
| **e** | MP | 0.0000 | 0.0000 | 0.0000 | 0.0000 | 0.0000 | 0.0000 | 0.0000 | 0.0000 | 0.0000 | 0.0000 | 0.0000 | 0.0000 | 0.0000 | 0.0000 | 0.0000 | 0.0000 |
|   | SF | 0.0000 | 0.2076 | 0.2113 | 0.2049 | 0.1935 | 0.1818 | 0.1698 | 0.1724 | 0.1776 | 0.2052 | 0.1737 | 0.1811 | 0.1836 | 0.1649 | 0.1726 | 0.1580 |
|   | SM | 0.0000 | 0.1267 | 0.1134 | 0.1190 | 0.1142 | 0.1140 | 0.1018 | 0.0993 | 0.1115 | 0.1179 | 0.1090 | 0.1025 | 0.1114 | 0.0957 | 0.0925 | 0.0974 |
| **f** | MP | 0.0000 | 0.0000 | 0.0000 | 0.0000 | 0.0000 | 0.0000 | 0.0000 | 0.0000 | 0.0000 | 0.0000 | 0.0000 | 0.0000 | 0.0000 | 0.0000 | 0.0000 | 0.0000 |
|   | SF | 0.0000 | 0.0166 | 0.0161 | 0.0175 | 0.0146 | 0.0139 | 0.0186 | 0.0182 | 0.0142 | 0.0200 | 0.0148 | 0.0151 | 0.0180 | 0.0178 | 0.0157 | 0.0130 |
|   | SM | 0.0000 | 0.0105 | 0.0103 | 0.0131 | 0.0091 | 0.0082 | 0.0120 | 0.0081 | 0.0102 | 0.0119 | 0.0102 | 0.0080 | 0.0071 | 0.0094 | 0.0102 | 0.0088 |

Note: We abbreviate our parental groups; namely, "MP" denotes married parents, "SF" denotes single fathers, and "SM" denotes single mothers. The leftmost column shows the corresponding category of relief eligibility for Scenario 1; the top row denotes the corresponding year.

**Table 2.1: Proportion of Parents Eligible for CTC Relief – S1**

| | | 2018 | 2017 | 2016 | 2015 | 2014 | 2013 | 2012 | 2011 | 2010 | 2009 | 2008 | 2007 | 2006 | 2005 | 2004 | 2003 |
|---|---|---|---|---|---|---|---|---|---|---|---|---|---|---|---|---|---|
| | MP | 0.0030 | 0.0016 | 0.0022 | 0.0016 | 0.0017 | 0.0016 | 0.0016 | 0.0017 | 0.0016 | 0.0017 | 0.0070 | 0.0178 | 0.0210 | 0.0131 | 0.0140 | 0.0152 |
| a | SF | 0.0059 | 0.0039 | 0.0031 | 0.0050 | 0.0021 | 0.0027 | 0.0043 | 0.0019 | 0.0058 | 0.0038 | 0.0144 | 0.0350 | 0.0394 | 0.0251 | 0.0268 | 0.0290 |
| | SM | 0.0098 | 0.0107 | 0.0079 | 0.0086 | 0.0081 | 0.0087 | 0.0095 | 0.0113 | 0.0089 | 0.0096 | 0.0399 | 0.0978 | 0.1013 | 0.0763 | 0.0741 | 0.0758 |
| | MP | 0.0395 | 0.0219 | 0.0218 | 0.0243 | 0.0245 | 0.0307 | 0.0285 | 0.0300 | 0.0325 | 0.0268 | 0.0418 | 0.0667 | 0.0559 | 0.0656 | 0.0708 | 0.0883 |
| b | SF | 0.0441 | 0.0293 | 0.0492 | 0.0453 | 0.0366 | 0.0470 | 0.0443 | 0.0493 | 0.0473 | 0.0390 | 0.0674 | 0.0478 | 0.0574 | 0.0658 | 0.0665 | 0.0651 |
| | SM | 0.1186 | 0.0972 | 0.1051 | 0.1051 | 0.1227 | 0.1156 | 0.1156 | 0.1199 | 0.1207 | 0.1219 | 0.1682 | 0.1147 | 0.1203 | 0.1437 | 0.1539 | 0.2068 |
| | MP | 0.3749 | 0.2795 | 0.2951 | 0.2937 | 0.2936 | 0.2692 | 0.2724 | 0.2720 | 0.2705 | 0.2614 | 0.2065 | 0.1908 | 0.1669 | 0.1755 | 0.1849 | 0.1373 |
| c | SF | 0.3071 | 0.1809 | 0.1836 | 0.1960 | 0.2087 | 0.2282 | 0.2043 | 0.2013 | 0.1834 | 0.1914 | 0.1492 | 0.1381 | 0.0958 | 0.1201 | 0.1134 | 0.1241 |
| | SM | 0.4378 | 0.3316 | 0.3435 | 0.3090 | 0.2988 | 0.3105 | 0.3157 | 0.3207 | 0.3081 | 0.3107 | 0.2069 | 0.2089 | 0.2062 | 0.2117 | 0.1668 | 0.1209 |
| | MP | 0.5826 | 0.6970 | 0.6809 | 0.6803 | 0.6801 | 0.6984 | 0.6975 | 0.6963 | 0.6955 | 0.7101 | 0.7448 | 0.7247 | 0.7562 | 0.7459 | 0.7303 | 0.7592 |
| d | SF | 0.6429 | 0.5285 | 0.5102 | 0.5005 | 0.5073 | 0.5022 | 0.5262 | 0.5291 | 0.5470 | 0.5109 | 0.5463 | 0.5566 | 0.5767 | 0.5803 | 0.5804 | 0.5828 |
| | SM | 0.4337 | 0.3983 | 0.3953 | 0.4274 | 0.4266 | 0.4237 | 0.4232 | 0.4204 | 0.4200 | 0.4092 | 0.4476 | 0.4491 | 0.4376 | 0.4428 | 0.4828 | 0.4752 |
| | MP | 0.0000 | 0.0000 | 0.0000 | 0.0000 | 0.0000 | 0.0000 | 0.0000 | 0.0000 | 0.0000 | 0.0000 | 0.0000 | 0.0000 | 0.0000 | 0.0000 | 0.0000 | 0.0000 |
| e | SF | 0.0000 | 0.2573 | 0.2539 | 0.2532 | 0.2452 | 0.2199 | 0.2209 | 0.2184 | 0.2165 | 0.2550 | 0.2226 | 0.2226 | 0.2307 | 0.2087 | 0.2129 | 0.1990 |
| | SM | 0.0000 | 0.1622 | 0.1482 | 0.1499 | 0.1439 | 0.1415 | 0.1360 | 0.1277 | 0.1423 | 0.1485 | 0.1374 | 0.1295 | 0.1346 | 0.1254 | 0.1224 | 0.1212 |
| | MP | 0.0000 | 0.0000 | 0.0000 | 0.0000 | 0.0000 | 0.0000 | 0.0000 | 0.0000 | 0.0000 | 0.0000 | 0.0000 | 0.0000 | 0.0000 | 0.0000 | 0.0000 | 0.0000 |
| f | SF | 0.0000 | 0.0000 | 0.0000 | 0.0000 | 0.0000 | 0.0000 | 0.0000 | 0.0000 | 0.0000 | 0.0000 | 0.0000 | 0.0000 | 0.0000 | 0.0000 | 0.0000 | 0.0000 |
| | SM | 0.0000 | 0.0000 | 0.0000 | 0.0000 | 0.0000 | 0.0000 | 0.0000 | 0.0000 | 0.0000 | 0.0000 | 0.0000 | 0.0000 | 0.0000 | 0.0000 | 0.0000 | 0.0000 |

Note: We abbreviate our parental groups; namely, "MP" denotes married parents, "SF" denotes single fathers, and "SM" denotes single mothers. The leftmost column shows the corresponding category of relief eligibility for Scenario 2; the top row denotes the corresponding year.

**Table 2.2: Proportion of Parents Eligible for CTC Relief – S2**

To account for a final potential source of error, we exclude households whose annual income exceeds $99,999, as the CPS Table Creator does not provide more specific income data beyond this threshold. Thus, we are unable to estimate whether parents who earn $100,000 or more qualify for given levels of CTC relief.

This exclusion impacts our estimates for Categories d to f of CTC relief eligibility. From 2003 – 2017, the phaseout threshold for married parents was $110,000 (8). Thus, we cannot determine whether and to what extent married parents earning $100,000 or more qualify for full CTC relief, some CTC relief, or no relief. Our figures for married parents' relief eligibility for Category d, are thereby slightly underestimated, as we are unable to account for parents who earn between $100,000 and $110,000 and thus would be eligible for full CTC relief. Owing to insufficient data granularity, for Category e we cannot provide an informed estimate, as households could fall into either (or neither) category depending on their specific income. For category f, the total number of households who fall above the total phaseout can be accurately estimated. However, those whose earnings exceed $99,999 are excluded from analysis to ensure homogeneity across scenarios for comparison purposes. For 2018, we also cannot provide an informed estimate for Categories, d, e, and f owing to insufficient specificity, as the phaseout threshold for married parents remained substantially above $99,999.

Using a similar approach, our estimates of single-parent relief eligibility for Categories d and e, which are based on phaseout thresholds from 2003 – 2017, are accurate, though our Category f figures are underestimated. For 2018, full CTC relief eligibility (i.e., Category d) for single parents is an underestimate. Owing to insufficient specificity, for Categories e and f, we cannot provide an informed estimate. As we cannot precisely estimate Category f for any parental group or time period, we focus on Categories a to e throughout most of our subsequent analysis.

**Section II**

Here, we analyze the income distribution of CPS respondents from 2003 – 2018 and highlight key differences across parental groups. We begin by leveraging our counts of married parents, single mothers, and single fathers broken down by income bracket (Section I) to estimate the proportion, rather than the number of parents, of each parental group whose income falls into each bin.

To do so, we employ a 2-step process. First, we sum the number of single fathers across all income brackets to calculate the total number of single fathers in a given year. We subsequently divide each count by the total to estimate for a given year what proportion of single fathers is contained within each income bracket. We then repeat this process year-on-year from 2003 – 2018 and for both single mothers and married parents. Our results are shown in Figures 1.1 and 1.2.

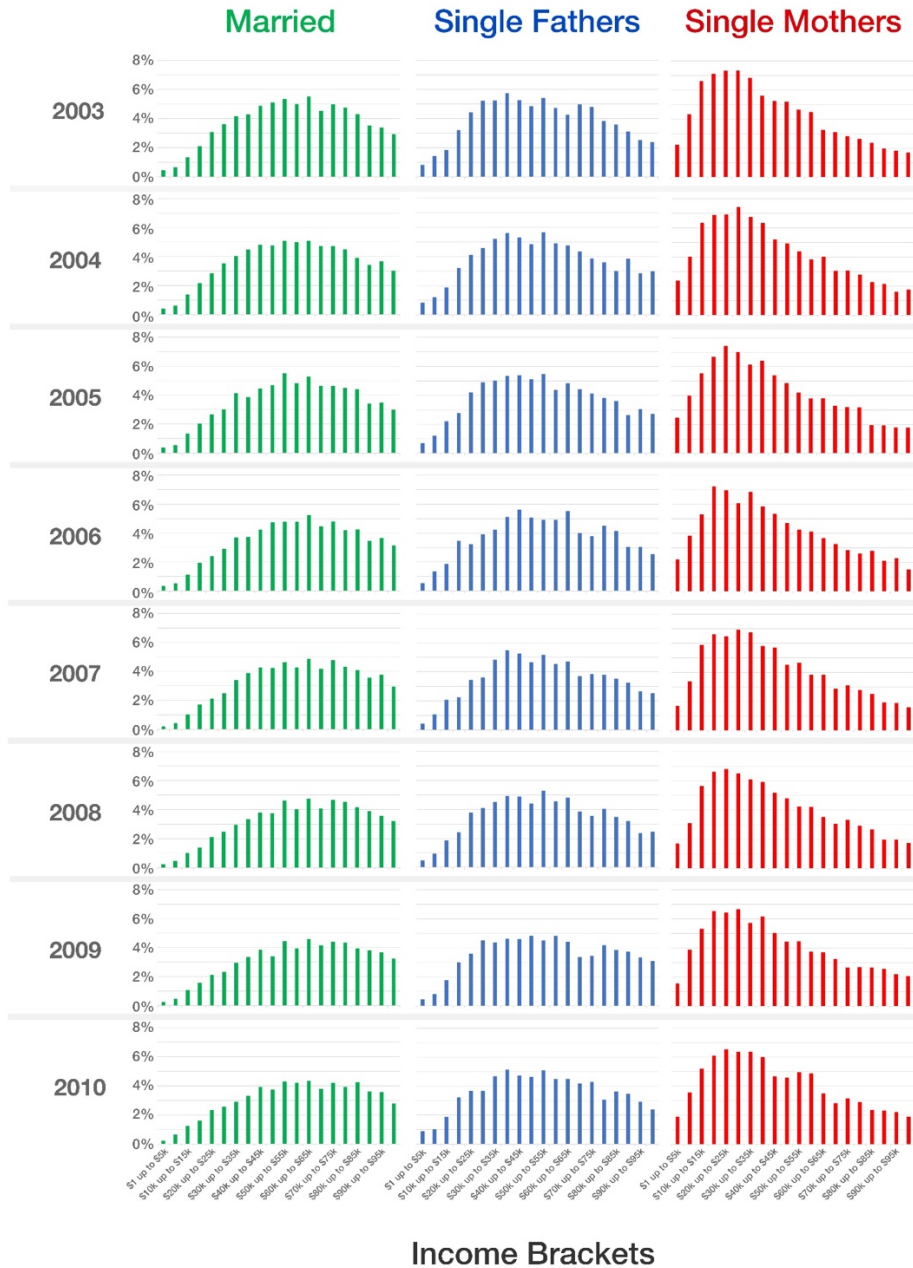

Figure 1.1: Income Distribution Analysis by Year (2003 – 2010)

Note: Each histogram represents a parental group's income distribution for a given year. Our figures are color-coded by parental group: green corresponds to married parents; blue corresponds to single fathers; and red corresponds to single mothers. The x-axis denotes the associated income bracket, measured in $5,000 increments from $1 to $99,999; the y-axis denotes the percentage of each parental group within each income bracket.

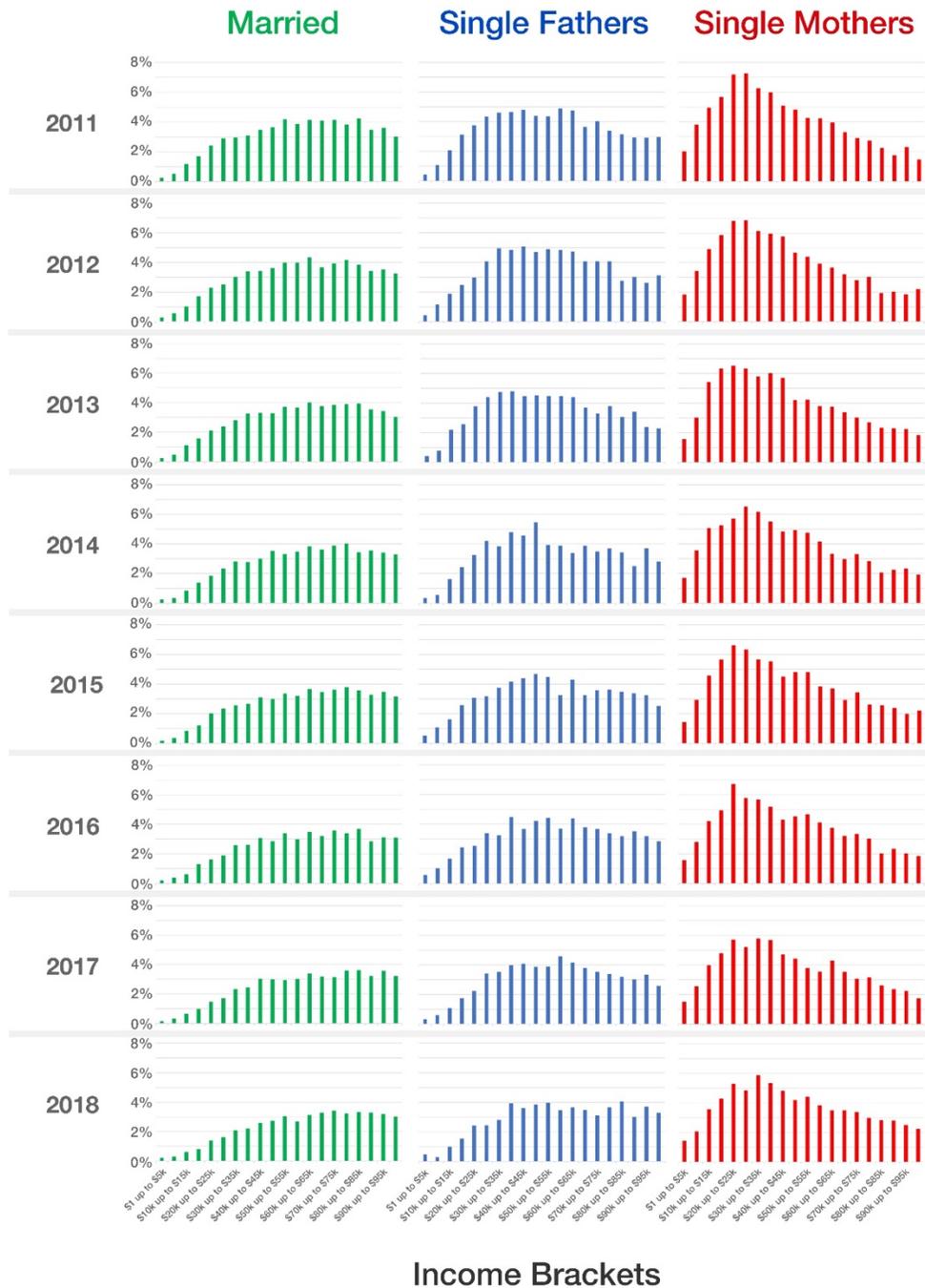

Note: Each histogram represents a parental group's income distribution for a given year. Our figures are color-coded by parental group: green corresponds to married parents; blue corresponds to single fathers; and red corresponds to single mothers. The x-axis denotes the associated income bracket, measured in $5,000 increments from $1 to $99,999; the y-axis denotes the percentage of each parental group within each income bracket.

**Figure 1.2: Income Distribution Analysis by Year (2011 – 2018)**

**Section III**

*Section 3.1: 2003 - 2017*

In this section, we detail our statistical and comparative analyses from 2003 to 2017. We first explore the impact of lowering the refundability threshold from $10,500 in 2003 to $3,000 by 2017. To investigate the interaction of gender and marital effects on increased CTC eligibility, we compared the proportions of single fathers who were ineligible for any relief owing to low income (Category a) in 2003 versus 2017. To estimate the benefit realized by single fathers as a result of the reduced refundability threshold, we calculated the difference between the proportion of ineligible single fathers (low income) before and after the change. We then repeated this process for single mothers and married parents.

Using a similar process, we also analyzed the impact of reducing the refundability threshold on parents' ability to claim full ACTC relief (Category c). To begin, we estimated the proportion of single fathers eligible for full ACTC relief in 2017. Next, we adjusted the refundability and max ACTC earnings thresholds to their 2003 levels and re-estimated this proportion. We then calculated the delta between our estimates to highlight the effect of changing the refundability threshold on single fathers and repeated this process for single mothers and married parents.

For each category of relief, we estimated a linear fixed effects regression from 2003 – 2017. Specifically, we regressed the proportion of parents, broken down by parental group and year-on-year, parental group, and interaction terms (i.e., year multiplied by parental group).

This allows us to analyze (1) whether programmatic changes to the CTC, occurring in particular years, differentially impacted single mothers or single fathers compared to married parents, and (2) whether there are any significant differences across the entire time period in the overall level of single mothers or single fathers qualifying for categories of relief as compared to married parents.

Our income analysis (Section II) suggests income requirements of the full CTC are more likely to impede mothers' eligibility than married parents' or single fathers'. This is reflected in our regression results (Table 3, Category d), as the coefficient for single mothers is statistically significant and negative, suggesting income requirements of full CTC disproportionately disadvantaged single mothers.

We also note that the coefficient for single fathers in Category d is statistically significant but smaller in magnitude than single mothers', providing further empirical evidence that it is beneficial (from a relief perspective) to be married for both men and women, but the effect is stronger for single mothers (Section 3.2, Main Paper).

Using our fixed effects model, we also assess differences in eligibility for some CTC relief (Category e) between single mothers and fathers. Our results suggest that single mothers are significantly less likely than single fathers to qualify for some CTC relief (Table 3, Category e).

| VARIABLES | Category a (S1) ineligible (low income) | Category b (S1) some actc | Category c (S1) fullactc | Category d (S1) fullctc | Category e (S1) somectc | Category f (S1) ineligible (high income) | Category a (S2) ineligible (low income) | Category b (S2) some actc | Category c (S2) fullactc | Category d (S2) fullctc | Category e (S2) somectc |
|---|---|---|---|---|---|---|---|---|---|---|---|
| SingleMen | 0.0020*** | 0.0037** | -0.0309*** | -0.1956*** | 0.2043*** | 0.0166*** | 0.0015 | 0.0095 | -0.0995*** | -0.1653*** | 0.2538*** |
|  | (0.0001) | (0.0019) | (0.0001) | (0.0177) | (0.0190) | (0.0003) | (0.0048) | (0.0114) | (0.0054) | (0.0189) | (0.0201) |
| SingleWomen | 0.0169*** | 0.0281*** | 0.0911*** | -0.2666*** | 0.1200*** | 0.0104*** | 0.0074*** | 0.0793*** | 0.0503*** | -0.2922*** | 0.1552*** |
|  | (0.0000) | (0.0000) | (0.0000) | (0.0000) | (0.0000) | (0.0000) | (0.0000) | (0.0000) | (0.0000) | (0.0000) | (0.0000) |
| Year2003 | 0.0202*** | 0.0319*** | -0.0552*** | 0.0065 | -0.0033 | -0.0001 | 0.0127*** | 0.0685*** | -0.1431*** | 0.0654*** | -0.0035 |
|  | (0.0001) | (0.0019) | (0.0001) | (0.0177) | (0.0190) | (0.0003) | (0.0048) | (0.0114) | (0.0054) | (0.0189) | (0.0201) |
| Year2004 | 0.0189*** | 0.0198*** | -0.0422*** | 0.0069 | -0.0033 | -0.0001 | 0.0116** | 0.0509*** | -0.0956*** | 0.0366* | -0.0035 |
|  | (0.0001) | (0.0019) | (0.0001) | (0.0177) | (0.0190) | (0.0003) | (0.0048) | (0.0114) | (0.0054) | (0.0189) | (0.0201) |
| Year2005 | 0.0185*** | 0.0323*** | -0.0653*** | 0.0179 | -0.0033 | -0.0001 | 0.0106** | 0.0457*** | -0.1049*** | 0.0521*** | -0.0035 |
|  | (0.0001) | (0.0019) | (0.0001) | (0.0177) | (0.0190) | (0.0003) | (0.0048) | (0.0114) | (0.0054) | (0.0189) | (0.0201) |
| Year2006 | 0.0177*** | 0.0304*** | -0.0367*** | -0.0081 | -0.0033 | -0.0001 | 0.0186*** | 0.0360*** | -0.1135*** | 0.0624*** | -0.0035 |
|  | (0.0001) | (0.0019) | (0.0001) | (0.0177) | (0.0190) | (0.0003) | (0.0048) | (0.0114) | (0.0054) | (0.0189) | (0.0201) |
| Year2007 | 0.0145*** | 0.0277*** | -0.0430*** | 0.0041 | -0.0033 | -0.0001 | 0.0153*** | 0.0469*** | -0.0897*** | 0.0310 | -0.0035 |
|  | (0.0001) | (0.0019) | (0.0001) | (0.0177) | (0.0190) | (0.0003) | (0.0048) | (0.0114) | (0.0054) | (0.0189) | (0.0201) |
| Year2008 | 0.0077*** | 0.0199*** | -0.0338*** | 0.0096 | -0.0033 | -0.0001 | 0.0045 | 0.0219* | -0.0740*** | 0.0510*** | -0.0035 |
|  | (0.0001) | (0.0019) | (0.0001) | (0.0177) | (0.0190) | (0.0003) | (0.0048) | (0.0114) | (0.0054) | (0.0189) | (0.0201) |
| Year2009 | 0.0005*** | 0.0010 | -0.0050*** | 0.0068 | -0.0033 | -0.0001 | -0.0007 | 0.0070 | -0.0191*** | 0.0163 | -0.0035 |
|  | (0.0001) | (0.0019) | (0.0001) | (0.0177) | (0.0190) | (0.0003) | (0.0048) | (0.0114) | (0.0054) | (0.0189) | (0.0201) |
| Year2010 | 0.0002 | 0.0041** | 0.0061*** | -0.0070 | -0.0033 | -0.0001 | -0.0009 | 0.0126 | -0.0099* | 0.0017 | -0.0035 |
|  | (0.0001) | (0.0019) | (0.0001) | (0.0177) | (0.0190) | (0.0003) | (0.0048) | (0.0114) | (0.0054) | (0.0189) | (0.0201) |
| Year2011 | 0.0009*** | 0.0018 | 0.0374*** | -0.0368** | -0.0033 | -0.0001 | -0.0008 | 0.0102 | -0.0085 | 0.0026 | -0.0035 |
| | (0.0001) | (0.0019) | (0.0001) | (0.0177) | (0.0190) | (0.0003) | (0.0048) | (0.0114) | (0.0054) | (0.0189) | (0.0201) |
| Year2012 | 0.0010*** | 0.0025 | 0.0295*** | -0.0297* | -0.0033 | -0.0001 | -0.0009 | 0.0087 | -0.0080 | 0.0037 | -0.0035 |
| | (0.0001) | (0.0019) | (0.0001) | (0.0177) | (0.0190) | (0.0003) | (0.0048) | (0.0114) | (0.0054) | (0.0189) | (0.0201) |
| Year2013 | 0.0009*** | 0.0020 | 0.0290*** | -0.0286 | -0.0033 | -0.0001 | -0.0009 | 0.0109 | -0.0112** | 0.0047 | -0.0035 |
| | (0.0001) | (0.0019) | (0.0001) | (0.0177) | (0.0190) | (0.0003) | (0.0048) | (0.0114) | (0.0054) | (0.0189) | (0.0201) |
| Year2014 | 0.0013*** | -0.0002 | 0.0225*** | -0.0202 | -0.0033 | -0.0001 | -0.0007 | 0.0047 | 0.0132** | -0.0136 | -0.0035 |
| | (0.0001) | (0.0019) | (0.0001) | (0.0177) | (0.0190) | (0.0003) | (0.0048) | (0.0114) | (0.0054) | (0.0189) | (0.0201) |
| Year2015 | 0.0000 | 0.0004 | 0.0238*** | -0.0208 | -0.0033 | -0.0001 | -0.0008 | 0.0045 | 0.0133** | -0.0135 | -0.0035 |
| | (0.0001) | (0.0019) | (0.0001) | (0.0177) | (0.0190) | (0.0003) | (0.0048) | (0.0114) | (0.0054) | (0.0189) | (0.0201) |
| Year2016 | 0.0006** | 0.0019 | 0.0153*** | -0.0111 | -0.0066 | -0.0001 | -0.0012 | 0.0040 | 0.0137 | -0.0096 | -0.0070 |
| | (0.0003) | (0.0037) | (0.0003) | (0.0355) | (0.0380) | (0.0007) | (0.0097) | (0.0228) | (0.0107) | (0.0378) | (0.0402) |
| Constant | 0.0033*** | 0.0064*** | 0.1457*** | 0.8412*** | 0.0033 | 0.0001 | 0.0025 | 0.0199* | 0.2804*** | 0.6937*** | 0.0035 |
| | (0.0001) | (0.0019) | (0.0001) | (0.0177) | (0.0190) | (0.0003) | (0.0048) | (0.0114) | (0.0054) | (0.0189) | (0.0201) |
| | | | | | | | | | | | |
| Observations | 45 | 45 | 45 | 45 | 45 | 45 | 45 | 45 | 45 | 45 | 45 |
| R-squared | 1.0000 | 1.0000 | 1.0000 | 0.9999 | 0.9998 | 1.0000 | 0.9999 | 0.9998 | 1.0000 | 0.9999 | 0.9999 |

Robust standard errors in parentheses

*** p<0.01, ** p<0.05, * p<0.1

Note: The coefficients of interest are those on *SingleMen* and *SingleWomen*. *SingleMen* is a dummy variable for single fathers, *SingleWomen* is a dummy variable for single mothers, and married parents are the baseline parental group. Our time variables are defined as *YearX*, where *X* indicates the year of each observation (e.g., *Year2009* is a dummy variable denoting observations that occurred in 2009). We omit the year 2017 from our regression to avoid perfect multicollinearity. We also include interaction terms between each year and parental group (e.g., *Year2008 * SingleMen*) in our regression but omit reporting coefficients and standard errors. Owing to insufficient data granularity, we cannot analyze category f of relief eligibility for S2.

**Table 3: Fixed Effects Regression Analysis**

Additionally, we analyze potential gender and marital effects owing to the parity that existed between full CTC and ACTC relief from 2009 to 2017. During this period, the maximum available relief from both sources was $1,000. To investigate the effect that such parity had on parental groups' relief eligibility, we follow a 3-step process. First, we calculate the total number of parents in each group who qualified for the full magnitude of CTC relief (Category c + d). Next, we estimate the total number of parents in each group who would be priced out of the full magnitude of relief if the maximum CTC were increased to $2,000 while maintaining a $1,000 maximum for the ACTC (see Section I for details). Finally, we find the proportion of parents in each group who are priced out by taking the total number of parents who are priced out divided by the total number of parents who qualified for full relief before raising the maximum CTC (see Tables 4.1 – 4.3). The proportion is derived using the following equation, repeated for each parental group:

$$proportion_{priced\_out} = \frac{priced\_out}{full\_relief_{old}} \quad (1)$$

where $priced_{out}$ = the raw count of the number of parents who fail to qualify for full relief once the maximum CTC is raised to $2,000, absent numerical parity; $full\_relief_{old}$ = the raw count of the number of parents who qualified for full relief under numerical parity between the CTC and ACTC; and $proportion_{priced\_out}$ = the ratio between $priced_{out}$ and $full\_relief_{old}$.

| | S1 | | | S2 | | |
|---|---|---|---|---|---|---|
| | **Raw Counts** | | **Proportion** | **Raw Counts** | | **Proportion** |
| | $full\_relief_{old}$ | $priced\_out$ | $proportion_{priced\_out}$ | $full\_relief_{old}$ | $priced\_out$ | $proportion_{priced\_out}$ |
| 2003 | 33029000 | 4063000 | 12.3 | 31317000 | 5376000 | 17.17 |
| 2004 | 32463000 | 4539000 | 13.98 | 31834000 | 7788000 | 24.46 |
| 2005 | 32110000 | 4158000 | 12.95 | 31498000 | 7231000 | 22.96 |
| 2006 | 31101000 | 4857000 | 15.62 | 30577000 | 6757000 | 22.1 |
| 2007 | 30228000 | 4419000 | 14.62 | 29749000 | 7622000 | 25.62 |
| 2008 | 28093000 | 3926000 | 13.98 | 27643000 | 6488000 | 23.47 |
| 2009 | 27796000 | 3956000 | 14.23 | 27309000 | 7348000 | 26.9 |
| 2010 | 26723000 | 4114000 | 15.39 | 26180000 | 7331000 | 28 |
| 2011 | 25817000 | 4788000 | 18.55 | 25314000 | 7110000 | 28.09 |
| 2012 | 24739000 | 4393000 | 17.76 | 24317000 | 6830000 | 28.09 |
| 2013 | 23995000 | 4247000 | 17.7 | 23516000 | 6543000 | 27.82 |
| 2014 | 23216000 | 3948000 | 17.01 | 22853000 | 6891000 | 30.15 |
| 2015 | 22486000 | 3851000 | 17.13 | 22124000 | 6672000 | 30.16 |
| 2016 | 21127000 | 3443000 | 16.3 | 20869000 | 6310000 | 30.24 |
| 2017 | 20841000 | 3067000 | 14.72 | 20558000 | 5884000 | 28.62 |
| **Average Proportion Priced Out (2003-2017)** | | | **15.48** | **Average Proportion Priced Out (2003-2017)** | | **26.26** |
| **Average Proportion Priced Out (2009-2017)** | | | **16.53** | **Average Proportion Priced Out (2009-2017)** | | **28.68** |

**Table 4.1: Proportion of Married Parents who are priced out of the full magnitude of relief, after the Max CTC is raised from $1000 to $2000, from 2009-2017**

| | S1 | | | S2 | | |
|---|---|---|---|---|---|---|
| | **Raw Counts** | | **Proportion** | **Raw Counts** | | **Proportion** |
| | $full\_relief_{old}$ | $priced\_out$ | $proportion_{priced\_out}$ | $full\_relief_{old}$ | $priced\_out$ | $proportion_{priced\_out}$ |
| 2003 | 4603000 | 582000 | 12.64 | 4326000 | 870000 | 20.11 |
| 2004 | 4512000 | 690000 | 15.29 | 4229000 | 893000 | 21.12 |
| 2005 | 4539000 | 695000 | 15.31 | 4223000 | 898000 | 21.26 |
| 2006 | 4470000 | 664000 | 14.85 | 4157000 | 822000 | 19.77 |
| 2007 | 4204000 | 591000 | 14.06 | 3904000 | 913000 | 23.39 |
| 2008 | 4236000 | 783000 | 18.48 | 3910000 | 985000 | 25.19 |
| 2009 | 4002000 | 821000 | 20.51 | 3713000 | 1012000 | 27.26 |
| 2010 | 3916000 | 742000 | 18.95 | 3660000 | 919000 | 25.11 |
| 2011 | 3768000 | 772000 | 20.49 | 3494000 | 963000 | 27.56 |
| 2012 | 3867000 | 653000 | 16.89 | 3579000 | 1001000 | 27.97 |
| 2013 | 3800000 | 755000 | 19.87 | 3527000 | 1102000 | 31.24 |
| 2014 | 3630000 | 691000 | 19.04 | 3341000 | 974000 | 29.15 |
| 2015 | 3475000 | 641000 | 18.45 | 3216000 | 905000 | 28.14 |
| 2016 | 3613000 | 630000 | 17.44 | 3356000 | 888000 | 26.46 |
| 2017 | 3472000 | 524000 | 15.09 | 3239000 | 826000 | 25.5 |
| **Average Proportion Priced Out (2003-2017)** | | | **17.16** | **Average Proportion Priced Out (2003-2017)** | | **25.28** |
| **Average Proportion Priced Out (2009-2017)** | | | **18.52** | **Average Proportion Priced Out (2009-2017)** | | **27.60** |

**Table 4.2: Proportion of Single Fathers who are priced out of the full magnitude of relief, after the Max CTC is raised from $1000 to $2000, from 2009-2017**

| | S1 | | | S2 | | |
|---|---|---|---|---|---|---|
| | **Raw Counts** | | **Proportion** | **Raw Counts** | | **Proportion** |
| | $full\_relief_{old}$ | $priced\_out$ | $proportion_{priced\_out}$ | $full\_relief_{old}$ | $priced\_out$ | $proportion_{priced\_out}$ |
| 2003 | 8119000 | 1825000 | 22.48 | 7457000 | 2255000 | 30.24 |
| 2004 | 8947000 | 2547000 | 28.47 | 7930000 | 2683000 | 33.83 |
| 2005 | 8816000 | 2491000 | 28.26 | 7895000 | 3126000 | 39.59 |
| 2006 | 8502000 | 2446000 | 28.77 | 7669000 | 3073000 | 40.07 |
| 2007 | 8719000 | 2416000 | 27.71 | 7769000 | 3041000 | 39.14 |
| 2008 | 8484000 | 2879000 | 33.93 | 7586000 | 2968000 | 39.12 |
| 2009 | 8095000 | 2712000 | 33.5 | 7252000 | 3130000 | 43.16 |
| 2010 | 7971000 | 2541000 | 31.88 | 7147000 | 3024000 | 42.31 |
| 2011 | 8015000 | 2552000 | 31.84 | 7230000 | 3129000 | 43.28 |
| 2012 | 8088000 | 2605000 | 32.21 | 7274000 | 3108000 | 42.73 |
| 2013 | 8046000 | 2634000 | 32.74 | 7197000 | 3044000 | 42.3 |
| 2014 | 7792000 | 2433000 | 31.22 | 6978000 | 2874000 | 41.19 |
| 2015 | 8037000 | 2584000 | 32.15 | 7286000 | 3057000 | 41.96 |
| 2016 | 7889000 | 2429000 | 30.79 | 7119000 | 3310000 | 46.5 |
| 2017 | 7461000 | 2186000 | 29.3 | 6737000 | 3061000 | 45.44 |
| **Average Proportion Priced Out (2003-2017)** | | | **30.35** | **Average Proportion Priced Out (2003-2017)** | | **40.72** |
| **Average Proportion Priced Out (2009-2017)** | | | **31.74** | **Average Proportion Priced Out (2009-2017)** | | **43.21** |

**Table 4.3: Proportion of Single Mothers who are priced out of the full magnitude of relief, after the Max CTC is raised from $1000 to $2000, from 2009-2017**

We then explore who benefits from the maximum available CTC relief irrespective of pathway (i.e., as a refund versus a credit). To do so, we combine our estimates for full relief eligibility via the ACTC and CTC into a single category (Category C + D) and estimate a linear fixed effects regression using our previously-discussed methodology. Our analysis suggests single mothers are overrepresented in full relief (see Table 5).

| VARIABLES | Full relief eligibility (S1) | Full relief eligibility (S2) |
|---|---|---|
| SingleMen | -0.2265*** | -0.2648*** |
| | (0.0176) | (0.0135) |
| SingleWomen | -0.1755*** | -0.2419*** |
| | (0.0000) | (0.0000) |
| Year2003 | -0.0487*** | -0.0777*** |
| | (0.0176) | (0.0135) |
| Year2004 | -0.0353** | -0.0590*** |
| | (0.0176) | (0.0135) |
| Year2005 | -0.0474*** | -0.0528*** |
| | (0.0176) | (0.0135) |
| Year2006 | -0.0448** | -0.0511*** |
| | (0.0176) | (0.0135) |
| Year2007 | -0.0388** | -0.0587*** |
| | (0.0176) | (0.0135) |
| Year2008 | -0.0242 | -0.0229* |
| | (0.0176) | (0.0135) |
| Year2009 | 0.0018 | -0.0027 |
| | (0.0176) | (0.0135) |
| Year2010 | -0.0009 | -0.0082 |
| | (0.0176) | (0.0135) |
| Year2011 | 0.0006 | -0.0059 |
| | (0.0176) | (0.0135) |
| Year2012 | -0.0002 | -0.0043 |
| | (0.0176) | (0.0135) |
| Year2013 | 0.0005 | -0.0065 |
| | (0.0176) | (0.0135) |
| Year2014 | 0.0023 | -0.0005 |
|  | (0.0176) | (0.0135) |
| Year2015 | 0.0031 | -0.0002 |
|  | (0.0176) | (0.0135) |
| Year2016 | 0.0042 | 0.0042 |
|  | (0.0352) | (0.0270) |
| Constant | 0.9869 | 0.9742*** |
|  | (0.0176) | (0.0135) |
|  |  |  |
| Observations | 45 | 45 |

Robust standard errors in parentheses
*** p<0.01, ** p<0.05, * p<0.1

Note: The coefficients of interest are those on *SingleMen* and *SingleWomen*. *SingleMen* is a dummy variable for single fathers, *SingleWomen* is a dummy variable for single mothers, and married parents are the baseline parental group. Our time variables are defined as *YearX*, where *X* indicates the year of each observation (e.g., *Year2009* is a dummy variable denoting observations that occurred in 2009). We omit the year 2017 from our regression to avoid perfect multicollinearity. We also include interaction terms between each year and parental group (e.g., *Year2008 * SingleMen*) in our regression but omit reporting coefficients and standard errors.

**Table 5: Full Relief Eligibility Fixed Effects Regression Analysis**

We subsequently analyze the extent to which refundability matters, and to whom, by assessing the proportion of each parental group eligible for any level of ACTC relief. To do so, we combine our estimates for full ACTC and some ACTC relief into a single category (Category B + C) and again estimate a linear fixed effects regression according to our methodology described above. Our analysis suggests that single mothers' overrepresentation in full relief eligibility is a potential consequence of their disproportionate representation in full ACTC relief eligibility (see Table 6).

| VARIABLES | Any ACTC relief eligibility (S1) | Any ACTC relief eligibility (S2) |
|---|---|---|
| SingleMen | -0.0273*** | -0.0900*** |
|  | (0.0017) | (0.0060) |
| SingleWomen | 0.1192*** | 0.1296*** |
|  | (0.0000) | (0.0000) |
| Year2003 | -0.0233*** | -0.0746*** |
|  | (0.0017) | (0.0060) |
| Year2004 | -0.0224*** | -0.0447*** |
|  | (0.0017) | (0.0060) |
| Year2005 | -0.0330*** | -0.0592*** |
|  | (0.0017) | (0.0060) |
| Year2006 | -0.0063*** | -0.0775*** |
|  | (0.0017) | (0.0060) |
| Year2007 | -0.0153*** | -0.0428*** |
|  | (0.0017) | (0.0060) |
| Year2008 | -0.0139*** | -0.0521*** |
|  | (0.0017) | (0.0060) |
| Year2009 | -0.0040** | -0.0121** |
|  | (0.0017) | (0.0060) |
| Year2010 | 0.0102*** | 0.0027 |
|  | (0.0017) | (0.0060) |
| Year2011 | 0.0393*** | 0.0017 |
|  | (0.0017) | (0.0060) |
| Year2012 | 0.0320*** | 0.0006 |
|  | (0.0017) | (0.0060) |
| Year2013 | 0.0310*** | -0.0003 |
|  | (0.0017) | (0.0060) |
| Year2014 | 0.0223*** | 0.0179*** |
|  | (0.0017) | (0.0060) |
| Year2015 | 0.0242*** | 0.0178*** |
|  | (0.0017) | (0.0060) |
| Year2016 | 0.0171*** | 0.0177 |
|  | (0.0034) | (0.0121) |
| Constant | 0.1521*** | 0.3003*** |
|  | (0.0017) | (0.0060) |
|  |  |  |
| Observations | 45 | 45 |
| R-squared | 1.0000 | 1.0000 |

Robust standard errors in parentheses

*** p<0.01, ** p<0.05, * p<0.1

Note: The coefficients of interest are those on *SingleMen* and *SingleWomen*. *SingleMen* is a dummy variable for single fathers, *SingleWomen* is a dummy variable for single mothers, and married parents are the baseline parental group. Our time variables are defined as *YearX*, where *X* indicates the year of each observation (e.g., *Year2009* is a dummy variable denoting observations that occurred in 2009). We omit the year 2017 from our regression to avoid perfect multicollinearity. We also include interaction terms between each year and parental group (e.g., *Year2008 * SingleMen*) in our regression but omit reporting coefficients and standard errors.

**Table 6: Any ACTC Relief Eligibility Fixed Effects Regression Analysis**

Due to the Tax Cuts and Jobs Act (TCJA), many parameters of the CTC were changed between 2017 and 2018; namely, maximum CTC, maximum ACTC, refundability threshold, phaseout thresholds, and standard deductions (9). In order to isolate the impact on relief eligibility of individual parameters amidst these simultaneous changes, we conduct a piecemeal analysis.

To begin, we calculate the percentage of each parental group qualifying for full ACTC and CTC under the 2017 CTC program's parameters. However, we use the 2018 income distribution rather than 2017 to account for potential effects owing to changes in exogenous factors (i.e., household income). After establishing this baseline, we then change individual parameters of the CTC program from 2017 to 2018 levels and estimate new percentages of each parental group who qualify for full CTC (see Section I). Specifically, we first adjust only the maximum CTC, followed by both maximum CTC and the standard deduction, and finally maximum CTC, the standard deduction, and phaseout thresholds.

We leverage a similar approach for full ACTC, changing individual programmatic parameters from 2017 to 2018 levels and estimating parental groups' eligibility for full ACTC following each change. We first adjust maximum ACTC, followed by both maximum ACTC and the refundability threshold.

We further evaluate the magnitude and significance of the TJCA's effects on parents' ability to qualify for full CTC and ACTC using a difference-in-differences analysis. Specifically, we assess whether the TCJA changes, which went into effect at the end of 2017, had a significant impact on the difference in proportions of single mothers versus single fathers qualifying for full ACTC, full CTC, and some CTC relief (Categories c – e).

To do so, we create an interaction term between our treatment variable (i.e., a dummy variable for single motherhood) and time variable (i.e., a dummy variable for whether a response occurred in 2018). Using this interaction term as our difference-in-difference estimator, we estimate our model over Categories c – e for S1 and S2.

Using our methodology described in Section I, we then assess the impact of a variety of potential programmatic changes and gender effects. First, we assess the overall effect of the TCJA on relief eligibility by comparing the proportion of parents eligible for any form of CTC relief in 2018 (Categories b – e) to the proportion of eligible parents in 2018 were they to be subjected to the 2003 CTC program's parameters.

We then estimate the effect of heterogeneity in the average number of children between single fathers and single mothers on full CTC relief eligibility (Category d). To do so, we first estimate the average proportion of each parental group that qualifies for full CTC relief from 2003 to 2017 for S1. Using these estimates, we next calculate the difference in average eligibility between single fathers and single mothers. We then re-estimate the income thresholds required to claim full CTC relief given each parental group's average number of dependents (Table 1) and, using these new income thresholds, we estimate full CTC eligibility by parental group given dependent heterogeneity. Subsequently, we again calculate the difference in average eligibility between single fathers and single mothers from 2003 to 2017. Finally, we compare the differences in full CTC relief eligibility with versus without accounting for dependent heterogeneity to estimate its impact on gender differences in relief realization.

Next, we estimate how restored parity between the full CTC and ACTC would affect existing gender differences in relief eligibility. Specifically, we calculate the proportion of single mothers and single fathers who would qualify for full CTC or ACTC relief if the ACTC were raised from $1,400 to $2,000 (thereby raising the requisite income threshold from $11,833 to $15,833) using our methodology from Section I. This subsequently reduces the gap between the percentage of single mothers and single fathers eligible for full relief from 18.28 (20.92) percent to 7.85 (11.68) percent (see Table 1c in main paper).

When the $2,000 benefit is realized as a combination as a credit and a refund, however, the minimum amount required to claim the full magnitude of relief is lower. Specifically, the requisite income threshold for full relief magnitude falls to $24,000 ($28,140) for single fathers and $24,000 ($28,500) for single mothers. Thus, the percent of single fathers and mothers able to claim the full magnitude of the credit rises. In particular, this proportion rises more for single mothers than single fathers, since more single mothers are represented low areas of the income distribution. Specifically, the gap between the percentage of single mothers and single fathers prior to the ACTC being raised from $1,400 to $2,000 falls from 18.28 (20.92) percent to 12.71 (13.56) percent.

Finally, we analyze the impact of increasing the maximum relief available via the CTC program, assuming parity between the ACTC and CTC. In particular, we estimate the proportion of each parental group eligible for full relief given maximum credit sizes ranging from $500 to $3,600 in 2017 and 2018 for both S1 and S2. We find that increasing the maximum credit size disproportionately affects single mothers more than married parents and single fathers, causing disproportionately more of them to fail to qualify for full relief (see Tables 8.1 and 8.2).

| Max. Credit ($) | | 500 | 600 | 700 | 800 | 900 | 1000 | 1100 | 1200 | 1300 | 1400 | 1500 | 1600 | 1700 | 1800 | 1900 | 2000 | 3000 | 3600 |
|---|---|---|---|---|---|---|---|---|---|---|---|---|---|---|---|---|---|---|---|
| **S1** | MC | 99.29 | 99.29 | 99.00 | 99.00 | 99.00 | 99.00 | 98.38 | 98.38 | 98.38 | 98.38 | 97.65 | 97.65 | 97.65 | 97.65 | 96.56 | 96.56 | 92.67 | 90.68 |
| | SF | 76.61 | 76.61 | 76.04 | 76.04 | 76.04 | 76.04 | 75.30 | 75.30 | 75.30 | 75.30 | 74.24 | 74.24 | 74.24 | 74.24 | 72.54 | 72.54 | 67.65 | 64.56 |
| | SM | 82.51 | 82.51 | 80.83 | 80.83 | 80.83 | 80.83 | 78.26 | 78.26 | 78.26 | 78.26 | 75.49 | 75.49 | 75.49 | 75.49 | 72.15 | 72.15 | 61.46 | 57.15 |
| **S2** | MC | 99.00 | 99.00 | 98.38 | 98.38 | 97.65 | 97.65 | 96.56 | 96.56 | 95.67 | 95.67 | 94.04 | 94.04 | 92.67 | 92.67 | 90.68 | 90.68 | 79.46 | 69.70 |
| | SF | 73.30 | 72.73 | 72.73 | 71.99 | 71.99 | 70.94 | 70.94 | 69.23 | 69.23 | 69.23 | 68.05 | 68.05 | 65.86 | 65.86 | 64.35 | 64.35 | 52.85 | 42.20 |
| | SM | 78.33 | 78.33 | 78.33 | 75.75 | 75.75 | 72.99 | 72.99 | 69.64 | 69.64 | 66.59 | 66.59 | 62.28 | 62.28 | 58.96 | 58.96 | 58.96 | 39.83 | 30.33 |

**Table 8.1: Proportion of each parental group qualifying for full relief given different maximum sizes in 2017**

| Max. Credit ($) | | 500 | 600 | 700 | 800 | 900 | 1000 | 1100 | 1200 | 1300 | 1400 | 1500 | 1600 | 1700 | 1800 | 1900 | 2000 | 3000 | 3600 |
|---|---|---|---|---|---|---|---|---|---|---|---|---|---|---|---|---|---|---|---|
| S1 | MC | 99.16 | 99.16 | 99.16 | 98.83 | 98.83 | 98.83 | 98.83 | 98.12 | 98.12 | 98.12 | 98.12 | 97.51 | 97.51 | 97.51 | 96.48 | 96.48 | 94.05 | 90.62 |
| | SF | 98.97 | 98.97 | 98.97 | 98.67 | 98.67 | 98.67 | 98.67 | 97.75 | 97.75 | 97.75 | 97.75 | 96.96 | 96.96 | 96.96 | 95.00 | 95.00 | 92.03 | 87.18 |
| | SM | 96.37 | 96.37 | 96.37 | 95.23 | 95.23 | 95.23 | 95.23 | 92.49 | 92.49 | 92.49 | 92.49 | 90.39 | 90.39 | 90.39 | 87.15 | 87.15 | 80.35 | 73.62 |
| S2 | MC | 98.83 | 98.83 | 98.12 | 98.12 | 97.51 | 97.51 | 96.48 | 96.48 | 95.75 | 95.75 | 94.05 | 94.05 | 92.74 | 92.74 | 90.62 | 90.62 | 79.97 | 70.81 |
| | SF | 98.97 | 98.67 | 98.67 | 98.67 | 97.75 | 96.96 | 96.96 | 96.96 | 95.00 | 95.00 | 94.30 | 94.30 | 92.03 | 92.03 | 90.09 | 90.09 | 76.72 | 70.69 |
| | SM | 96.37 | 95.23 | 95.23 | 95.23 | 92.49 | 90.39 | 90.39 | 87.15 | 87.15 | 84.57 | 84.57 | 84.57 | 80.35 | 80.35 | 77.38 | 77.38 | 58.44 | 49.05 |

**Table 8.2: Proportion of each parental group qualifying for full relief given different maximum sizes in 2018**

**References**

1. "Current Population Survey, Annual Social and Economic Supplement." *US Census Bureau*, 2018.

2. "Annual Social and Economic Supplement (ASEC) of the Current Population Survey (CPS)." *US Census Bureau*, 8 Oct. 2021, https://www.census.gov/programs-surveys/saipe/guidance/model-input-data/cpsasec.html.

3. "Subject Definitions." *US Census Bureau*, 8 Oct. 2021, https://www.census.gov/programs-surveys/cps/technical-documentation/subject-definitions.html#:~:text=Own%20children%20in%20a%20family,or%20parent%20in%20the%20subfamily.

4. "Concepts and Definitions (CPS)." *U.S. Bureau of Labor Statistics*, 21 Oct. 2021, https://www.bls.gov/cps/definitions.htm.

5. "Tax Code, Regulations and Official Guidance." *Internal Revenue Service*, 13 Jan. 2022, https://www.irs.gov/privacy-disclosure/tax-code-regulations-and-official-guidance.

6. Orem, Tina. "How Do I Choose the Right Tax Filing Status?" *NerdWallet*, 6 Jan. 2022, https://www.nerdwallet.com/article/taxes/how-to-choose-tax-filing-status.

7. "The Child Tax Credit: Legislative History." *Congressional Research Service*, 1 Mar. 2018.

8. LaJoie, Taylor. "The Child Tax Credit: Primer." *Tax Foundation*, 14 Apr. 2020, https://taxfoundation.org/child-tax-credit/#:~:text=As%20part%20of%20the%20Tax,to%20%24400%2C000%20for%20married%20couples.

9. "How Did the TCJA Change Taxes of Families with Children?" *Tax Policy Center*, Urban Institute & Brookings Institution, May 2020, https://www.taxpolicycenter.org/briefing-book/how-did-tcja-change-taxes-families-children.

## Appendix

| | | 2003 | 2004 | 2005 | 2006 | 2007 | 2008 | 2009 | 2010 | 2011 | 2012 | 2013 | 2014 | 2015 | 2016 | 2017 | 2018 |
|---|---|---|---|---|---|---|---|---|---|---|---|---|---|---|---|---|---|
| Envisioned benefit | Married | 1,000 | 1,000 | 1,000 | 1,000 | 1,000 | 1,000 | 1,000 | 1,000 | 1,000 | 1,000 | 1,000 | 1,000 | 1,000 | 1,000 | 1,000 | 2,000 |
| | Single | 1,000 | 1,000 | 1,000 | 1,000 | 1,000 | 1,000 | 1,000 | 1,000 | 1,000 | 1,000 | 1,000 | 1,000 | 1,000 | 1,000 | 1,000 | 2,000 |
| Realizable benefit | Married | 1,000 | 1,000 | 1,000 | 1,000 | 1,000 | 1,000 | 1,000 | 1,000 | 1,000 | 1,000 | 1,000 | 1,000 | 1,000 | 1,000 | 1,000 | 2,000 |
| (credit only pathway) | | 28,650 | 29,000 | 29,600 | 30,200 | 30,900 | 31,400 | 32,350 | 32,350 | 32,700 | 33,300 | 33,900 | 34,250 | 34,600 | 34,750 | 34,850 | 43,850 |
| | Single | 1,000 | 1,000 | 1,000 | 1,000 | 1,000 | 1,000 | 1,000 | 1,000 | 1,000 | 1,000 | 1,000 | 1,000 | 1,000 | 1,000 | 1,000 | 2,000 |
| | | 23,100 | 23,350 | 23,700 | 24,150 | 24,650 | 25,000 | 25,650 | 25,700 | 25,900 | 26,300 | 26,750 | 27,000 | 27,250 | 27,400 | 27,450 | 36,950 |
| Realizable benefit | Married | 1,000 | 1,000 | 1,000 | 1,000 | 1,000 | 1,000 | 1,000 | 1,000 | 1,000 | 1,000 | 1,000 | 1,000 | 1,000 | 1,000 | 1,000 | 1,400 |
| (credit and/or refund | | 19,565 | 17,414 | 17,664 | 17,964 | 18,414 | 15,164 | 9,664 | 9,664 | 9,664 | 9,664 | 9,664 | 9,664 | 9,664 | 9,664 | 9,664 | 11,830 |
| pathway) | Single | 1,000 | 1,000 | 1,000 | 1,000 | 1,000 | 1,000 | 1,000 | 1,000 | 1,000 | 1,000 | 1,000 | 1,000 | 1,000 | 1,000 | 1,000 | 1,400 |
| | | 16,795 | 15,787 | 16,070 | 16,437 | 16,907 | 15,097 | 9,664 | 9,664 | 9,664 | 9,664 | 9,664 | 9,664 | 9,664 | 9,664 | 9,664 | 11,830 |
| Credit and/or Refund | Married | 907 | 1,000 | 1,000 | 1,000 | 1,000 | 1,000 | 1,000 | 1,000 | 1,000 | 1,000 | 1,000 | 1,000 | 1,000 | 1,000 | 1,000 | 1,400 |
| Breakdown | | 93 | - | - | - | - | - | - | - | - | - | - | - | - | - | - | - |
| | Single | 630 | 756 | 761 | 771 | 774 | 990 | 1,000 | 1,000 | 1,000 | 1,000 | 1,000 | 1,000 | 1,000 | 1,000 | 1,000 | 1,400 |
| | | 370 | 244 | 239 | 229 | 226 | 10 | - | - | - | - | - | - | - | - | - | - |

Note: "Envisioned benefit" indicates the magnitude of available benefits for parents qualifying for Full CTC relief (Category D) (expressed in $). "Realizable benefit" denotes the maximum available benefit on the upper line and the requisite income to realize such benefits on the lower line. "Credit and/or Refund Breakdown" describes the amount of relief realized as a credit versus a refund, owing to specificities in the pre-2009 US tax code. – denotes values that are not applicable.

**Table A1a: Maximum available magnitude of relief for each parental group, year on year (S1).**

| | | 2003 | 2004 | 2005 | 2006 | 2007 | 2008 | 2009 | 2010 | 2011 | 2012 | 2013 | 2014 | 2015 | 2016 | 2017 | 2018 |
|---|---|---|---|---|---|---|---|---|---|---|---|---|---|---|---|---|---|
| Standard Deduction | Married | 9,500 | 9,700 | 10,000 | 10,300 | 10,700 | 10,900 | 11,400 | 11,400 | 11,600 | 11,900 | 12,200 | 12,400 | 12,600 | 12,600 | 12,700 | 24,000 |
| | Single | 7,000 | 7,150 | 7,300 | 7,550 | 7,850 | 8,000 | 8,350 | 8,400 | 8,500 | 8,700 | 8,950 | 9,100 | 9,250 | 9,300 | 9,350 | 18,000 |
| Personal Exemption | Married | 9,150 | 9,300 | 9,600 | 9,900 | 10,200 | 10,500 | 10,950 | 10,950 | 11,100 | 11,400 | 11,700 | 11,850 | 12,000 | 12,150 | 12,150 | - |
| | Single | 6,100 | 6,200 | 6,400 | 6,600 | 6,800 | 7,000 | 7,300 | 7,300 | 7,400 | 7,600 | 7,800 | 7,900 | 8,000 | 8,100 | 8,100 | - |
| Refundability Threshold | Married | 10,500 | 10,750 | 11,000 | 11,300 | 11,750 | 8,500 | 3,000 | 3,000 | 3,000 | 3,000 | 3,000 | 3,000 | 3,000 | 3,000 | 3,000 | 2,500 |
| | Single | 10,500 | 10,750 | 11,000 | 11,300 | 11,750 | 8,500 | 3,000 | 3,000 | 3,000 | 3,000 | 3,000 | 3,000 | 3,000 | 3,000 | 3,000 | 2,500 |
| Phase out | Married | 110,000 | 110,000 | 110,000 | 110,000 | 110,000 | 110,000 | 110,000 | 110,000 | 110,000 | 110,000 | 110,000 | 110,000 | 110,000 | 110,000 | 110,000 | 400,000 |
| | Single | 75,000 | 75,000 | 75,000 | 75,000 | 75,000 | 75,000 | 75,000 | 75,000 | 75,000 | 75,000 | 75,000 | 75,000 | 75,000 | 75,000 | 75,000 | 200,000 |
| Total Phase out | Married | 130,000 | 130,000 | 130,000 | 130,000 | 130,000 | 130,000 | 130,000 | 130,000 | 130,000 | 130,000 | 130,000 | 130,000 | 130,000 | 130,000 | 130,000 | 440,000 |
| | Single | 95,000 | 95,000 | 95,000 | 95,000 | 95,000 | 95,000 | 95,000 | 95,000 | 95,000 | 95,000 | 95,000 | 95,000 | 95,000 | 95,000 | 95,000 | 240,000 |

Note: – denotes values that are not applicable.

**Table A1b: Applicable CTC program parameters year on year by parental group (S1) (expressed in $).**

| | | 2003 | 2004 | 2005 | 2006 | 2007 | 2008 | 2009 | 2010 | 2011 | 2012 | 2013 | 2014 | 2015 | 2016 | 2017 | 2018 |
|---|---|---|---|---|---|---|---|---|---|---|---|---|---|---|---|---|---|
| Envisioned benefit | Married | 1,890 | 1,900 | 1,900 | 1,890 | 1,890 | 1,900 | 1,890 | 1,890 | 1,880 | 1,890 | 1,890 | 1,880 | 1,900 | 1,890 | 1,890 | 3,780 |
| | Single Father | 1,680 | 1,710 | 1,710 | 1,680 | 1,680 | 1,690 | 1,670 | 1,680 | 1,670 | 1,670 | 1,690 | 1,710 | 1,690 | 1,700 | 1,660 | 3,380 |
| | Single Mother | 1,720 | 1,750 | 1,760 | 1,740 | 1,730 | 1,750 | 1,740 | 1,740 | 1,710 | 1,730 | 1,740 | 1,730 | 1,730 | 1,740 | 1,740 | 3,500 |
| Realizable benefit (credit only pathway) | Married | 1,890 | 1,900 | 1,900 | 1,890 | 1,890 | 1,900 | 1,890 | 1,890 | 1,880 | 1,890 | 1,890 | 1,880 | 1,900 | 1,890 | 1,890 | 3,780 |
| | | 38,631 | 39,223 | 40,013 | 40,770 | 41,743 | 42,567 | 43,765 | 43,782 | 44,156 | 45,082 | 45,921 | 46,309 | 47,017 | 47,138 | 47,271 | 58,675 |
| | Single Father | 1,680 | 1,710 | 1,710 | 1,680 | 1,680 | 1,690 | 1,670 | 1,680 | 1,670 | 1,670 | 1,690 | 1,710 | 1,690 | 1,700 | 1,660 | 3,380 |
| | | 29,707 | 30,351 | 30,855 | 31,177 | 31,895 | 32,498 | 33,212 | 33,365 | 33,562 | 34,113 | 34,958 | 35,521 | 35,660 | 35,985 | 35,640 | 48,433 |
| | Single Mother | 1,720 | 1,750 | 1,760 | 1,740 | 1,730 | 1,750 | 1,740 | 1,740 | 1,710 | 1,730 | 1,740 | 1,730 | 1,730 | 1,740 | 1,740 | 3,500 |
| | | 30,096 | 30,742 | 31,349 | 31,775 | 32,399 | 33,108 | 33,934 | 33,984 | 33,977 | 34,741 | 35,486 | 35,734 | 36,087 | 36,414 | 36,497 | 49,433 |
| Realizable benefit (credit and/or refund pathway) | Married | 1,890 | 1,900 | 1,900 | 1,890 | 1,890 | 1,900 | 1,890 | 1,890 | 1,880 | 1,890 | 1,890 | 1,880 | 1,900 | 1,890 | 1,890 | 2,646 |
| | | 25,380 | 22,765 | 23,190 | 23595 | 24,180 | 21,167 | 15,600 | 15,600 | 15,533 | 15,600 | 15,600 | 15,533 | 15,667 | 15,600 | 15,600 | 20,140 |
| | Single Father | 1,680 | 1,710 | 1,710 | 1,680 | 1,680 | 1,690 | 1,670 | 1,680 | 1,670 | 1,670 | 1,690 | 1,710 | 1,690 | 1,700 | 1,660 | 2,366 |
| | | 21,235 | 19,510 | 19,830 | 20,057 | 20,555 | 18,825 | 14,133 | 14,200 | 14,133 | 14,133 | 14,267 | 14,400 | 14,267 | 14,333 | 14,067 | 18,165 |
| | Single Mother | 1,720 | 1,750 | 1,760 | 1,740 | 1,730 | 1,750 | 1,740 | 1,740 | 1,710 | 1,730 | 1,740 | 1,730 | 1,730 | 1,740 | 1,740 | 2,450 |
| | | 21,500 | 19,720 | 20,093 | 20,375 | 20,823 | 19,150 | 14,600 | 14,600 | 14,400 | 14,533 | 14,600 | 14,533 | 14,533 | 14,600 | 14,600 | 18,500 |
| Credit and/or Refund Breakdown | Married | 1,488 | 1,802 | 1,829 | 1844 | 1,865 | 1,900 | 1,890 | 1,890 | 1,880 | 1,890 | 1,890 | 1,880 | 1,900 | 1,890 | 1,890 | 2,646 |
| | | 402 | 98 | 71 | 46 | 25 | - | - | - | - | - | - | - | - | - | - | - |
| | Single Father | 1,074 | 1,314 | 1,325 | 1,314 | 1,321 | 1,549 | 1,670 | 1,680 | 1,670 | 1,670 | 1,690 | 1,710 | 1,690 | 1,700 | 1,660 | 2,350 |
| | | 606 | 396 | 386 | 366 | 359 | 141 | - | - | - | - | - | - | - | - | - | 17 |
| | Single Mother | 1,100 | 1,346 | 1,364 | 1,361 | 1,361 | 1,598 | 1,740 | 1,740 | 1,710 | 1,730 | 1,740 | 1,730 | 1,730 | 1,740 | 1,740 | 2,400 |
| | | 620 | 405 | 396 | 378 | 369 | 153 | - | - | - | - | - | - | - | - | - | 50 |

Note: "Envisioned benefit" indicates the magnitude of available benefits for parents qualifying for Full CTC relief (Category D) (expressed in $). "Realizable benefit" denotes the maximum available benefit on the upper line and the requisite income to realize such benefits on the lower line. "Credit and/or Refund Breakdown" describes the amount of relief realized as a credit versus a refund, owing to specificities in the pre-2009 US tax code. – denotes values that are not applicable.

**Table A2a: Maximum available magnitude of relief for each parental group, year on year (S2).**

| | | 2003 | 2004 | 2005 | 2006 | 2007 | 2008 | 2009 | 2010 | 2011 | 2012 | 2013 | 2014 | 2015 | 2016 | 2017 | 2018 |
|---|---|---|---|---|---|---|---|---|---|---|---|---|---|---|---|---|---|
| Standard Deduction | Married | 9,500 | 9,700 | 10,000 | 10,300 | 10,700 | 10,900 | 11,400 | 11,400 | 11,600 | 11,900 | 12,200 | 12,400 | 12,600 | 12,600 | 12,700 | 24,000 |
| | Single Father | 7,000 | 7,150 | 7,300 | 7,550 | 7,850 | 8,000 | 8,350 | 8,400 | 8,500 | 8,700 | 8,950 | 9,100 | 9,250 | 9,300 | 9,350 | 18,000 |
| | Single Mother | 7,000 | 7,150 | 7,300 | 7,550 | 7,850 | 8,000 | 8,350 | 8,400 | 8,500 | 8,700 | 8,950 | 9,100 | 9,250 | 9,300 | 9,350 | 18,000 |
| Personal Exemption | Married | 11,865 | 12,090 | 12,480 | 12,837 | 13,226 | 13,650 | 14,199 | 14,199 | 14,356 | 14,782 | 15,171 | 15,326 | 15,600 | 15,755 | 15,755 | - |
| | Single Father | 8,174 | 8,401 | 8,672 | 8,844 | 9,112 | 9,415 | 9,746 | 9,782 | 9,879 | 10,146 | 10,491 | 10,705 | 10,760 | 10,935 | 10,773 | - |
| | Single Mother | 8,296 | 8,525 | 8,832 | 9,042 | 9,282 | 9,625 | 10,001 | 10,001 | 10,027 | 10,374 | 10,686 | 10,784 | 10,920 | 11,097 | 11,097 | - |
| Refundability | Married | 10,500 | 10,750 | 11,000 | 11,300 | 11,750 | 8,500 | 3,000 | 3,000 | 3,000 | 3,000 | 3,000 | 3,000 | 3,000 | 3,000 | 3,000 | 2,500 |
| Threshold | Single Father | 10,500 | 10,750 | 11,000 | 11,300 | 11,750 | 8,500 | 3,000 | 3,000 | 3,000 | 3,000 | 3,000 | 3,000 | 3,000 | 3,000 | 3,000 | 2,500 |
| | Single Mother | 10,500 | 10,750 | 11,000 | 11,300 | 11,750 | 8,500 | 3,000 | 3,000 | 3,000 | 3,000 | 3,000 | 3,000 | 3,000 | 3,000 | 3,000 | 2,500 |
| Phase out | Married | 110,000 | 110,000 | 110,000 | 110,000 | 110,000 | 110,000 | 110,000 | 110,000 | 110,000 | 110,000 | 110,000 | 110,000 | 110,000 | 110,000 | 110,000 | 400,000 |
| | Single Father | 75,000 | 75,000 | 75,000 | 75,000 | 75,000 | 75,000 | 75,000 | 75,000 | 75,000 | 75,000 | 75,000 | 75,000 | 75,000 | 75,000 | 75,000 | 200,000 |
| | Single Mother | 75,000 | 75,000 | 75,000 | 75,000 | 75,000 | 75,000 | 75,000 | 75,000 | 75,000 | 75,000 | 75,000 | 75,000 | 75,000 | 75,000 | 75,000 | 200,000 |
| Total Phase out | Married | 147,800 | 148,000 | 148,000 | 147,800 | 147,800 | 148,000 | 147,800 | 147,800 | 147,600 | 147,800 | 147,800 | 147,600 | 148,000 | 147,800 | 147,800 | 475,600 |
| | Single Father | 108,600 | 109,200 | 109,200 | 108,600 | 108,600 | 108,800 | 108,400 | 108,600 | 108,400 | 108,400 | 108,800 | 109,200 | 108,800 | 109,000 | 108,200 | 267,600 |
| | Single Mother | 109,400 | 110,000 | 110,200 | 109,800 | 109,600 | 110,000 | 109,800 | 109,800 | 109,200 | 109,600 | 109,600 | 109,600 | 109,600 | 109,800 | 109,800 | 270,000 |

Note: – denotes values that are not applicable.

Table A2b: Applicable CTC program parameters year on year by parental group (S2) (expressed in $).

| | Relief Categories | | | | | |
|---|---|---|---|---|---|---|
| | **Ineligible for relief (low income) (a)** | **Eligible for some ACTC (b)** | **Eligible for full ACTC (c)** | **Eligible for full CTC (d)** | **Eligible for some CTC (e)** | **Ineligible for relief (high income) (f)** |
| **Married Parents (pre 2018)** | Accurate | Accurate | Accurate | Underestimate $32,350 (35,599) * to $110,000 | Unavailable $110,000 to $130,000 | Unavailable $130,000+ |
| **Married Parents (2018)** | Accurate | Accurate | Accurate | Unavailable $43,850 to $400,000 | Unavailable $400,000 to $440,000 | Unavailable $440,000+ |
| **Single Parents (pre 2018)** | Accurate | Accurate | Accurate | Accurate | Accurate | Underestimate[+] $95,000+ |
| **Single Parents (2018)** | Accurate | Accurate | Accurate | Underestimate $36,950 to $200,000 | Unavailable $200,000 to $240,000 | Unavailable $240,000+ |

\* Median requisite full CTC income threshold from 2003 – 2017

[+] This figure is deliberately underestimated to ensure comparability across our analyses of all scenarios and periods.

Note: The upper line denotes congruence with general population, the lower line denotes the income / phase out thresholds that inform congruence assessment.

**Table A3a: Generalizability of relief eligibility estimates (S1).**

| | Relief Categories | | | | | |
|---|---|---|---|---|---|---|
| | Ineligible for relief (low income) (a) | Eligible for some ACTC (b) | Eligible for full ACTC (c) | Eligible for full CTC (d) | Eligible for some CTC (e) | Ineligible for relief (high income) (f) |
| **Married Parents (pre 2018)** | Accurate | Accurate | Accurate | Underestimate $43781.8* to $110,000 | Unavailable $110,000 to $147800* | Unavailable $147800+* |
| **Married Parents (2018)** | Accurate | Accurate | Accurate | Unavailable $58675 to $400,000 | Unavailable $400,000 to $475,600 | Unavailable $475,600+ |
| **Single Parents (pre 2018)** | Accurate | Accurate | Accurate | Accurate | Underestimate $75,000 to $108600 (Single Fathers)*/$109800 (Single Mothers) | Unavailable $108600+* (Single Fathers)/$109800+* (Single Mothers) |
| **Single Parents (2018)** | Accurate | Accurate | Accurate | Underestimate $48433.3 (Single Fathers)/$49433.3 (Single Mothers) to $200,000 | Unavailable $200,000 to $267600 (Single Fathers)/$270000 (Single Mothers) | Unavailable $267600+ (Single Fathers)/$270000+ (Single Mothers) |

Note: The upper line denotes congruence with general population, the lower line denotes the income / phase out thresholds that inform congruence assessment.

**Table A3b: Generalizability of relief eligibility estimates (S2).**